\UseRawInputEncoding
 \documentclass[final,5p,times,twocolumn,authoryear]{elsarticle}

\usepackage{amssymb}
\usepackage{amsmath}
\usepackage{subcaption}
\usepackage{algorithm}
\usepackage{algorithmicx}
\usepackage[noend]{algpseudocode}
\usepackage{tabularx}
\usepackage[hyphens]{url}
\usepackage{hyperref} 
\hypersetup{
  colorlinks   = true,
  breaklinks   = true
}

\newcommand{\arcsec}{''}

\journal{Astronomy $\&$ Computing}

\begin{document}
\begin{frontmatter}

\title{Survey-Wide Asteroid Discovery with a High-Performance Computing Enabled Non-Linear Digital Tracking Framework}

\author[1]{Nathan Golovich}\ead{golovich1@llnl.gov}
\author[1]{Trevor Steil}\ead{steil1@llnl.gov}
\author[1]{Alex Geringer-Sameth}
\author[1]{Keita Iwabuchi}
\author[2]{Ryan Dozier}
\author[1]{Roger Pearce}
\affiliation[1]{organization={Lawrence Livermore National Laboratory},
                addressline={7000 East Avenue},
                city={Livermore},
                postcode={94550},
                state={California},
                country={United States of America}}
\affiliation[2]{organization={University of Central Florida},
                addressline={4328 Scorpius Street},
                city={Orlando},
                postcode={32816},
                state={Florida},
                country={United States of America}}

\begin{abstract}
Modern astronomical surveys detect asteroids by linking together their appearances across multiple images taken over time. This approach faces limitations in detecting faint asteroids and handling the computational complexity of trajectory linking. We present a novel method that adapts ``digital tracking" - traditionally used for short-term linear asteroid motion across images - to work with large-scale synoptic surveys such as the Vera Rubin Observatory Legacy Survey of Space and Time (Rubin/LSST). Our approach combines hundreds of sparse observations of individual asteroids across their non-linear orbital paths to enhance detection sensitivity by several magnitudes. To address the computational challenges of processing massive data sets and dense orbital phase spaces, we developed a specialized high-performance computing architecture. We demonstrate the effectiveness of our method through experiments that take advantage of the extensive computational resources at Lawrence Livermore National Laboratory. This work enables the detection of significantly fainter asteroids in existing and future survey data, potentially increasing the observable asteroid population by orders of magnitude across different orbital families, from near-Earth objects (NEOs) to Kuiper belt objects (KBOs).\end{abstract}

\end{frontmatter}

\section{Introduction}
\label{introduction}

Astronomical surveys that repeatedly image the sky with wide-field-of-view imagers offer the best chance to find minor planets. Astronomers typically detect moving objects in individual images and link them together across consecutive images on the basis of plausible trajectories. This method faces two primary challenges. First, the Solar System contains many minor planets, making it computationally expensive to link the correct detections without confusion, especially in very sensitive surveys. Second, since most minor planets are intrinsically very faint, astronomers must detect them on multiple images during the limited time spans when they are close to Earth. Exposures cannot be arbitrarily long or sensitivity decreases due to trailing losses. To address these challenges, surveys use a strategy that allows astronomers to make and link repeat detections over minutes to days with enough precision for follow-up observations at other observatories. 

Today, there are a handful of surveys that detect most asteroids, including the Catalina Sky Survey~\citep[CSS;][]{CSS}, the Panoramic Survey Telescope and Rapid Response System~\citep[Pan-STARRS;][]{panstarrs}, and the Asteroid Terrestrial-impact Last Alert System~\citep[ATLAS;][]{atlas}. Each of these surveys are geared toward detecting near-Earth objects (NEOs), which are minor planets (i.e., asteroids and comets) with perihelion less than 1.3 astronomical units (AU). NEOs are the target for several dedicated surveys because their detection is the vital first step for planetary defense. 

Other populations of minor planets remain further away from Earth, and thus pose no direct risk, but their study is scientifically useful for understanding Solar System formation and evolution and, indirectly, for planetary defense since all NEOs are sourced from further out in the Solar System before they enter orbits that bring them closer to Earth. These populations range from the main belt asteroids (MBAs) to the Kuiper belt objects (KBOs) and have been surveyed far less systematically across the Solar System. These objects are inherently fainter because of their larger distances, so longer exposures are required to detect them. However, all known objects move appreciably over time. For example, Sedna, among the farthest of the known minor planets, moves at $\sim 0.3\arcsec\,\text{day}^{-1}$. 

In general, the farther out an object is, the longer the time period over which its apparent motion can be approximated as linear. Motions may be approximated as linear across the sky for as long as a few weeks for KBOs and as long as a night for MBAs~\citep[see Figure~4 of][]{Heinze15}. This approximately linear motion has been exploited to enable a simplified stacking procedure along parallel linear trajectories known colloquially as ``digital tracking'', ``synthetic tracking'' or ``track-before-detect''. The method has been used for about 30 years to study trans-Neptunian objects~\citep[TNOs;][]{Tyson92,Cochran95, Bernstein04,WhiddenGPU}, MBAs~\citep{Heinze15,Heinze19}, and NEOs~\citep{Shao14, Zhai14, Zhai18, Zhai20, Lifset21}. Each of these examples of digital tracking methodologies have similar survey strategies: a spatially small survey composed of a dense stack of images spanning a short period of time during which linear motion is a valid approximation. The signal-to-noise ratio (SNR) for image stacking scales as $\sqrt{N}$, where $N$ is the number of exposures. Detection limits can therefore be pushed fainter by a few magnitudes for stacks of hundreds of exposures.

Over the next decade, the most prolific minor planet discovery engine will be the Vera C. Rubin Observatory Legacy Survey of Space and Time~\citep[LSST;][]{ivezic2008lsst}, which will observe the entire available night sky every few nights with six optical filters. Data will be automatically processed in real time, including the generation of difference images and automatic asteroid alerts (among other optical transients) as well as an automated asteroid detection and linking procedure~\citep{LSSTDM,Heliolinc}. The large aperture of the Rubin Observatory will deliver an unprecedented look into the orbital and luminosity distributions for all types of minor planets. The expected number of detected objects is $5.5\times10^6$ MBAs, $10^{5}$ NEOs, $2.8\times10^{5}$ Jovian Trojans, and $4\times10^{4}$ TNOs and KBOs~\citep[see Figure~5.1 of][]{abell2009lsst}. The typical asteroid will be detected hundreds of times over ten years~\citep[see Figure~5.4 of][]{abell2009lsst}.

This situation raises the question of whether or not these repeat observations can be used to enhance the sensitivity of LSST to even fainter solar system bodies. The basic idea is to view the hundreds of intersections between a minor planet's orbit and LSST images similarly to the concept of digital tracking described above. The key difference is that digital tracking has typically only been used over short time spans\footnote{\citet{Bernstein04} carried out a non-linear search for TNOs in stacks of 55 images taken over 5 days with the Hubble Space Telescope's ACS camera.} or those over which linear motion is a valid approximation. The ability to stack over years of imaging would translate to orders of magnitude more detections in surveys like LSST. 

\begin{figure}
	\centering 
\includegraphics[width=0.85\columnwidth]{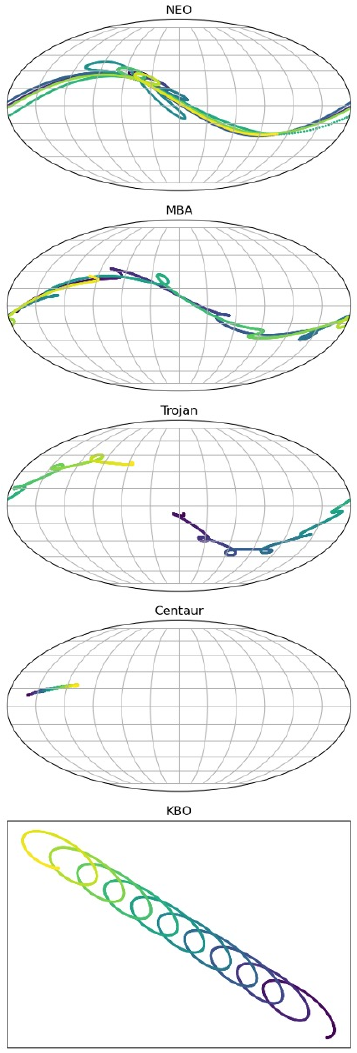}
	\caption{Projected sky motion over 10 years for minor planets from various populations ranging from the inner to outer Solar System. The color indicates time. For the KBO example, the object only moves only $\sim0.15\deg$ over ten years. The helical looping is a parallax effect caused by Earth's yearly motion around the sun while the broader trajectory is due to the minor planet's motion.} 
	\label{fig:skymotion}
\end{figure}

The challenge in doing so is computational. First, the generalization from linear motion to sky-projected orbital motion is complicated. Figure~\ref{fig:skymotion} shows the on-sky motion for various types of minor planets. Each of these curves is parametrized by six orbital elements and an orbital epoch. Geringer-Sameth et al. (in preparation) develops a general methodology to quantify density in the space of orbital elements as projected onto sky image data. They show that non-linear digital tracking searches are feasible for the outer Solar System with large high performance computing (HPC) systems. For linear digital tracking, an analogous calculation is presented in \S2.3 of~\citet{Heinze15}. Systematic searches for asteroids in the inner solar system are much more challenging with non-linear digital tracking. First, perturbations from the planets cause deviations from Keplerian motion over years-long surveys and must be accounted for. Second, the shorter distances mean that extremely small changes in orbital elements lead to detectable differences on the sky. The takeaway is that digital tracking in the non-linear regime should, at present, only be considered for minor planets beyond Jupiter. 

Whether considering digital tracking in the linear or non-linear regime, the method is only as powerful as the number of individual image epochs that can be combined. Since the linear regime is bounded in the time domain, this forces experiments to gather data that are spatially narrow and dense in the time domain. All-sky synoptic surveys (such as LSST and the Zwicky Transient Facility, ZTF) are fundamentally different in that the subsets of the data that are spatially narrow are necessarily sparse in the time domain~\citep[with the exception of the so-called deep drilling fields planned for LSST;][]{abell2009lsst}.

Combining the concept of digital tracking with synoptic surveys requires a novel computing architecture that can handle the sparse data intersections, large overall data volume, and extremely high required orbital phase space density. The features of such an architecture map well onto a number of computer science applications already deployed on HPC architectures, some of which have been explicitly developed to exploit the HPC architectures common at our place of work, Lawrence Livermore National Laboratory (LLNL). LLNL has long been home to cutting-edge HPC resources, and at the time of writing, the world's highest-performing HPC system is newly commissioned at LLNL~\citep{el-capitan}.

This paper describes the digital tracking framework that we have built for asteroid detection using LLNL HPC systems. This work is able to take advantage of the massive parallelism available through distributed graphics processing unit (GPU) systems to perform digital tracking of large numbers of orbits through telescope surveys with millions of images. The use of GPUs and HPC is relatively rare in astronomy, so in this paper we detail the scaling and performance of our methodology. The remainder of this paper is organized as follows. In \S\ref{sec:data} we discuss the data that we use to develop and test our method. In \S\ref{sec:framework} we discuss the computing framework that we assume for our method and how it influences the design of our algorithm. In \S\ref{sec:results} we present experiments using our algorithm to demonstrate performance. Finally in \S\ref{sec:discussion} we summarize our results and discuss the future development of our pipeline.

\section{Data}\label{sec:data}
LLNL was an institutional member of phase two of the Zwicky Transient Facility~\citep{ztf}. ZTF is an optical time-domain survey that uses the Palomar 48-inch Schmidt telescope. It has a 47~deg$^{2}$ field of view and a 600~megapixel camera with an eight-second read-out time. The survey is capable of imaging the northern sky at 4000~deg$^{2}$~hour$^{-1}$ to median depths of $g\sim20.8$ and $r\sim20.6$ (AB, $5\sigma$ in 30~seconds). The camera is composed of 16 charge-coupled devices (CCDs), each with $\sim6000^2$ pixels. Each CCD outputs four quadrant files, which means that every image taken results in 64 files. For more details on the ZTF data reduction pipeline, we refer to~\citet{ztf2}. 

LLNL has a copy of all the $g$, $r$ and $i$-band difference images produced by ZTF from March 2018 to February 2024, totaling $\sim400$~TB on disk and $\sim56$~million CCD quadrant files. Data are stored as \texttt{fpack} compressed FITS files~\citep{fits,fpack}. ZTF data are available at \url{https://irsa.ipac.caltech.edu/Missions/ztf.html}.

\section{Computing Framework}\label{sec:framework}

We designed our asteroid detection pipeline to leverage the power of modern heterogeneous computing architectures, using the computational power of GPUs to intersect orbits with images, the storage capacity of node-local solid-state drives (SSDs) to provide fast access to massive partitioned data sets, and the low latency of HPC networks to coordinate between processes running on the system. Pseudocode for this pipeline is given in Algorithm~\ref{algorithm:asteroid_search}.

\begin{algorithm*}[t!]
    \caption{Asteroid Search}
    \label{algorithm:asteroid_search}
    \begin{algorithmic}[1]
        \Require{$image\_list$, $batch\_size$, $num\_batches$}
        \Ensure{$DetectedObjects$}
        \State $local\_image\_list \gets$ PartitionImages($image\_list$) \Comment{Assign images to processes}
        \State $local\_images, local\_image\_headers \gets$ LoadImages($local\_image\_list$) \Comment{Read assigned images and extract header information}
        \State $DetectedObjects \gets \emptyset$
        \For{$i \gets 1$ to $num\_batches$}
            \State $orbits \gets$ GenerateOrbits($batch\_size$)
            \State $gpu\_orbits \gets$ CopyOrbitsToGPU($orbits$)
            \State $gpu\_intersections \gets$ SearchImages($gpu\_orbits, local\_image\_headers$) \Comment{Dense $|local\_images| \times |orbits|$ intersections}
            \State $filtered\_intersections \gets$ FilterGPUIntersections($gpu\_intersections$) \Comment{Filter invalid intersections on GPU}
            \State $intersections \gets$ CopyIntersectionsToHost($gpu\_intersections$)
            \State $local\_results \gets$ FindLocalSignal($intersections, local\_images$)
            \State $combined\_results \gets$ CommunicateResults($local\_results$)
            \State $detections \gets$ FindSignificantResults($combined\_results$)
            \State $DetectedObjects \gets DetectedObjects \cup detections$
        \EndFor
        \Return $DetectedObjects$
    \end{algorithmic}
\end{algorithm*}

This framework specifically targets the Lassen supercomputer at LLNL.\footnote{\url{https://hpc.llnl.gov/hardware/compute-platforms/lassen}} Lassen contains 795 compute nodes, each of which features two IBM Power9 central processing units (CPUs) totaling 44 cores, four NVidia V100 GPUs each with 7.8 TFLOP/s of double precision performance and 16 GB of second-generation high bandwidth memory (HBM2), 256 GB of random access memory (RAM), a 1.2 TB SSD, and a dual-port Mellanox 100 Gb/s Infiniband network card. Although our work specifically targets Lassen, we have chosen to use libraries (discussed in \S~\ref{sec:portability}) to make our code portable to new systems as they become available.

Our system starts by storing the relevant images for a given experiment on the SSDs available on each compute node. In order to maximize the number of images that can be handled, they are evenly partitioned across processes running on the set of available compute nodes without replication. When searching for asteroids, the CPUs sample sets of orbital parameters from specified distributions, with all CPUs being synchronized to sample the same orbits in the same order. These orbits are then transferred in batches to the GPUs. The \texttt{SearchImages} function of Algorithm~\ref{algorithm:asteroid_search} takes these batches of Keplerian orbital elements, identifies images that contain an object on each orbit, and computes the pixel coordinates of the objects within the intersected images. Orbits assume elliptical two-body motion around the Sun,\footnote{The Sun's motion model is inertial with velocity and initial position set by the DE440 model at the midpoint of ZTF.} with the location of the Earth at the time of every image predicted from the JPL DE440 solar system model~\citep{2021AJ....161..105P}, and ZTF's location computed with \texttt{astropy}\footnote{Specifically, \texttt{astropy.coordinates.EarthLocation.get\_gcrs\_posvel}, using ZTF's geocentric location listed by the Minor Planet Center}~\citep{astropy:2022}. Transformation from sky coordinates to image coordinates is done according to the World Coordinate System~\citep{2002A&A...395.1077C} projection specified in the ZTF image headers.  which handle the non-linear tracking to determine which images are intersected by each orbit, and where within the relevant images those intersections occur. This step is a brute-force check of each $(image, \, orbit)$ pair for intersections. We partition this computation across GPUs so that each GPU only considers intersections with the images that have been assigned to its compute node.

After we have found the set of intersections for the current batch of orbits, we filter the results to only contain those that are within the boundaries of an image. We transfer these compacted results from the GPU to the CPU, and for each valid intersection returned, a patch of pixels is pulled from the associated image on the node's SSD. In \texttt{FindLocalSignal} in Algorithm~\ref{algorithm:asteroid_search}, detection significance is calculated with a matched filter assuming a Gaussian point spread function whose width is given in the image header and pixel noise variance calculated with an iterative version of the median absolute deviation on each image’s pixels. Signal from multiple images is combined assuming the asteroid has the same flux in each image, though a future a version will incorporate zero-point magnitude and a light curve based on the Sun-asteroid-Earth geometry~\citep{1989aste.conf..524B}.

Each orbit is assigned a process that is responsible for assembling the results from individual processes running across the system to determine if an asteroid is present on the orbit. This communication of results is performed asynchronously by having processes send their valid results to the appropriate process as they are found.

\begin{figure}[t!]
	\centering 
	\includegraphics[width=\columnwidth]{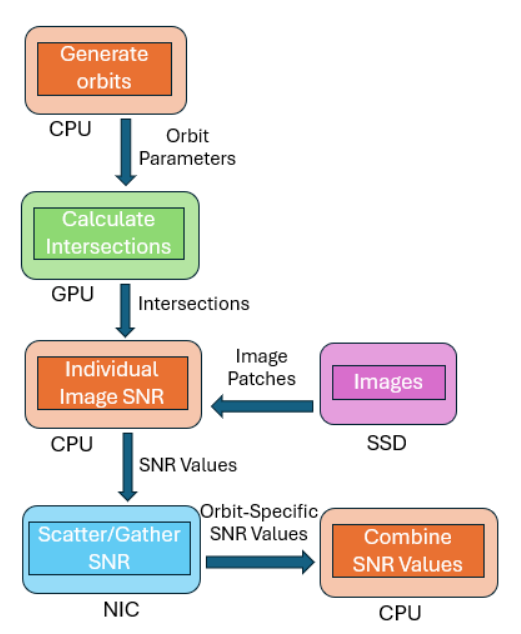}	
	\caption{Flow of data within the asteroid detection pipeline. The CPU is largely in charge of coordinating the computation by sampling orbits that the GPU uses for the dense intersection calculations, pulling the appropriate portions of images from the node-local SSDs, and communicating individual-image signal-to-noise results (SNR) with processes on compute nodes across the system to determine if a detection was made.} 
	\label{fig:data_flow}%
\end{figure}

The flow of data during asteroid detection is shown in Figure~\ref{fig:data_flow}. The GPU does the most FLOP intensive computations in order to determine where orbits intersect with images, taking the orbit parameters provided by the CPU and returning the locations of intersections. The CPU generates orbits used by the GPU, coordinates the collection of image data from the SSD for local image calculations, and sends its local results across the network to be combined with the results from other compute nodes in the system. The movement of data is all within a compute node except for the scatter and gather of the signal-to-noise ratio (SNR) results from the images local to each compute node to the node assigned to the orbit. The assigned node makes the final determination of whether a candidate orbit contained a detection by jointly combining the signal from all intersections.

We have begun the process of making our code openly available. It will appear on the LLNL GitHub\footnote{\url{https://github.com/LLNL}} page after the internal review and the release is complete.

\subsection{Computational Challenges}
\label{sec:computational_challenges}

Searching for minor planets in data from telescope surveys presents many challenges, starting with the sheer amount of data that needs to be accessed, the limited amount of memory available on GPUs, the need to communicate sparse and irregular local results from individual processes to determine if a detection was made with a given set of orbital elements, and the desire for portable and performant code that can be utilized across HPC systems. The computational framework is guided by the constraints posed by the combination of these computational challenges.

\subsubsection{Data Volume}
ZTF generates data at a rate of approximately 300~TB per year. In order to perform a meaningful survey of a population of possible asteroids, a significant portion of these images must be accessed. By holding these images on a parallel file system and accessing them as necessary, the rate at which images can be retrieved would be prohibitively slow for this approach. To alleviate this bottleneck, we instead stage collections of images necessary for a family of orbits on the local SSDs of our systems, giving us significantly faster random access to individual images.

To improve read performance, we tile images in smaller two-dimensional blocks as they are read from the parallel file system to each compute node's SSD. This layout makes the pixels we need to gather from an image more likely to be on a small number of memory pages, reducing the bandwidth necessary to read a patch of an individual image. If images are stored in a row or column-wise fashion on a system using 4~KB pages, only 1024 pixel values can be stored on a single page when using 4~byte floating point types. As this is less than the width or height of the images we are using, any small patch of pixels we need to retrieve will have all of the values within a single row on a single page (or sometimes two), but values from distinct rows will be on separate pages. Typically, we are retrieving a square patch of pixels across many rows and columns, so many pages would be necessary to read when not tiling images. By tiling our image into squares of 32 pixels by 32 pixels and assuming we are interested in a 10 pixel by 10 pixel square, the patch of pixels would be contained in a single page approximately half of the time, and the worst case for any individual patch is being split across 4 pages. This method reduces necessary bandwidth to the SSD by a factor of approximately five on average.

Additionally, we \emph{memory map} images on the node-local SSDs using the file-backed memory mapping technology (\texttt{mmap(2)} system call).\footnote{see \url{https://man7.org/linux/man-pages/man2/mmap.2.html}} By memory mapping, each image file is directly mapped to a region of the virtual memory space of the process and can be accessed as if it were stored on the main memory (DRAM) directly. This increases image data reading performance (higher bandwidth and lower latency) from the SSD by caching data on DRAM. This technique also has the benefit of delegating the complex data cache and transfer management from the application to the operating system~\citep{vanessen2012di-mmap}.

\subsubsection{GPU Memory}
\label{sec:gpu_memory}
Testing large numbers of candidate orbits against a massive telescope survey data set requires large amounts of computing power, making this repeated calculation ideal for offloading onto a GPU. Each intersection operation requires converting the set of orbital parameters into a sky position at the time an image was taken, followed by converting these celestial coordinates to locations within the pixel grid of an image. The total memory requirements of running this computation on the GPU can be computed as
\begin{equation}
    M_{GPU} = n_i m_i + n_o m_{o} + n_i n_o m_{i o},
    \label{eq:gpu_mem}
\end{equation}
where $M_{GPU}$ is the total GPU memory required, $n_i$ is the number of images with metadata on the GPU, $n_o$ is the number of orbits being processed in a batch, $m_{i}$ is the amount of memory needed for the metadata of an image, $m_{o}$ is the memory required for orbital parameters, and $m_{i o}$ is the memory needed to define an intersection between an image and an orbit. In our setting $m_{i}$ is hundreds of bytes, and $m_{o}$ and $m_{i o}$ are tens of bytes each.

If we were to load the image metadata for a survey with one million images on a single GPU, we would need hundreds of megabytes of memory for the metadata itself, but then we would also need tens of megabytes for all the intersection data from each orbit. This would allow us to process at most hundreds of orbits on a single GPU with 16 gigabytes of memory available, before we even account for intermediate data. This problem is made worse by the fact that we are sharing GPU resources among processes, meaning this hundreds of orbits number would need to be across all processes sharing a GPU.

Requiring such small batches of orbits would limit performance by forcing frequent synchronizations between the host and GPU. To avoid this cost, we choose to partition the image metadata across our set of GPUs using the same partitioning scheme as the images on SSDs. This lets us increase our value for $n_o$ by reducing the value of $n_i$ seen on each of our GPUs.

As an alternative, a larger value of $n_o$ could be used by streaming the image metadata to the GPU. This design would likely lead to the often-seen bottlenecks of transferring data to GPUs~\citep{mohan2020DNN_data_stalls,leclerc2023ffcv_bottlenecks}. By partitioning image metadata across GPUs, the only data transferred between the CPU and GPU is the orbital parameters describing a batch of orbits and the data about intersections. The intersection data is initially large, as it is the $n_i n_o m_{io}$ term in Equation~\ref{eq:gpu_mem}, but we are able to filter out obvious non-intersections before transferring data back to the CPU. Because of its advantages in terms of allowing larger batches of orbits to be processed at one time and its low data transfer needs between the CPU and GPU, we choose to partition image metadata in our work.

\subsubsection{Communication of Results}
\label{sec:communication}
After computing the intersection locations for a batch of orbits in the images partitioned throughout the system, each compute node holds a collection of partial results necessary to make a detection on individual orbits. These partial results held by an individual process are extremely sparse, as the probability is very low for a random orbit to intersect any of the thousands of images a particular process is responsible for.

The sparsity of results leads to an ``all-to-many'' communication pattern. Each process has results to communicate and receive, and an individual process is only going to communicate with a small but non-trivial subset of processes in the system. In order to handle these irregular communication patterns, we make use of YGM~\citep{steil2023ygm}, an asynchronous communication library designed for these types of all-to-many patterns in HPC. In this setting, processes independently assemble the local results from the images available and then asynchronously send these results to their destination. At the destination, the local results are assembled into a global assessment of a particular orbit once all local results have been received. This asynchronous method of computing allows individual processes to begin combining the partial results from other processes as they become available without having to wait for all other processes to finish communication before starting.

\subsection{Performance Portability}
\label{sec:portability}
When writing scientific software, runtimes and programming models can be utilized to achieve high performance on available HPC systems. These techniques often tie a project to specific hardware. The true desire is the ability to run scientific software on a variety of systems without maintaining versions of code for every architecture being targeted, while also achieving close to optimal performance. 

Multiple performance portability solutions are available to make these conflicting goals more attainable. Kokkos is a C++ performance portability library~\citep{edwards2014kokkos} that uses the abstraction of execution spaces and memory spaces to specify the physical hardware on which the code will run and where a piece of data resides on the various memory devices available to a system. RAJA~\citep{beckingsale2019raja} is a similar performance portability library that abstracts computation from the architecture but allows users to more explicitly control details such as memory management and memory access patterns.

In our case, we want to run on different systems based on the amount of available RAM and SSD space while making use of any accelerators present for the floating point-intensive intersection calculations. We chose to use the RAJA performance portability library for its ability to express loop-level parallelism that can be run on CPUs and multiple accelerators, while being easy to integrate with existing code and libraries. This allows us to write computationally intensive portions of code using RAJA to be offloaded to accelerators while being free to use YGM to handle the communication in our algorithm, as described in \S~\ref{sec:communication}.

\section{Experimental Results}
\label{sec:results}

For testing of our HPC asteroid detection pipeline, we use LLNL's Lassen supercomputer described in \S~\ref{sec:framework}. As a system, Lassen showcases the heterogeneous architecture with performant GPUs for compute-intensive orbit intersection calculations, node-local SSDs for storing large image data sets, and a high-bandwidth network for coordination of the results spread across each processor in the system.

\subsection{Verification Studies}
\label{sec:verification}

In order to verify that our HPC asteroid detection pipeline is producing correct results, we perform verification studies comparing to results obtained from an asteroid detection pipeline separately developed in Python. This Python implementation uses \texttt{SSAPy}\footnote{\url{https://github.com/LLNL/SSAPy}}~\citep{yeager2023cislunar,ssapy} to propagate orbits in time and \texttt{astropy}\footnote{\texttt{astropy}'s WCS implementation is based on \texttt{WCSLIB}~\citep{wcslib}.}~\citep{astropy:2013, astropy:2018, astropy:2022} to determine the pixel coordinates of intersections between orbits and images. It has been extensively tested against JPL Horizons\footnote{\label{horizons}\url{https://ssd.jpl.nasa.gov/horizons}}~\citep{2011jsrs.conf...87G} and bright asteroids in ZTF. We used the Python implementation to verify the HPC pipeline's predicted location of intersections and final detection SNR.

\subsubsection{Intersection Location Verification}

\begin{figure}[h!]
    \centering
    \includegraphics[width=\columnwidth]{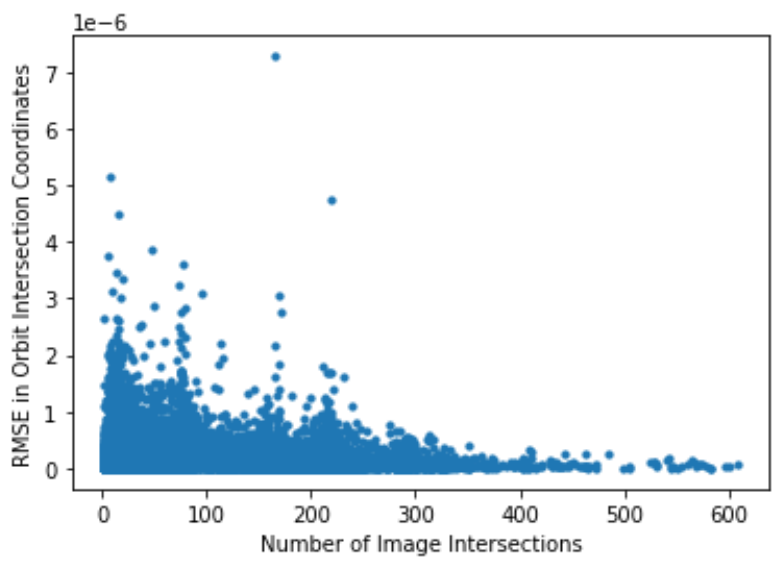}
    \caption{$\mathrm{RMSE}$ for asteroid pixel coordinates within intersected images compared to calculations using \texttt{SSAPy} and \texttt{astropy}.}
    \label{fig:orbit_rmse}
\end{figure}

For intersection verification of our HPC detection pipeline, we compare with the Python implementation across a collection of sampled orbits. For this experiment, we tested a collection of $\sim25,000$ randomly sampled orbits against two million images from the ZTF data set. Of these orbits, $\sim90$\% intersected at least one of the images used, yielding $\sim2$ million total intersections. We compare our implementations using the root mean square error ($\mathrm{RMSE}$) of intersections in pixels for an individual orbit, calculated as
\begin{equation*}
    \mathrm{RMSE} = \sqrt{\frac{1}{N}\sum_{i=1}^N \|X_i - \hat{X}_i\|_{2}^2}
\end{equation*}
where the $X_i$ are the intersection locations in pixel space as given by Python and the $\hat{X}_i$ are the intersection locations determined by the HPC pipeline. Figure~\ref{fig:orbit_rmse} shows the $\mathrm{RMSE}$ for each orbit, separated by the number of intersections given by each orbit. We can see most error values clustered around zero, with the largest errors approaching $10^{-5}$ pixels. 

\subsubsection{Detection Verification}

\begin{figure}[h!]
    \centering
    \includegraphics[width=\columnwidth]{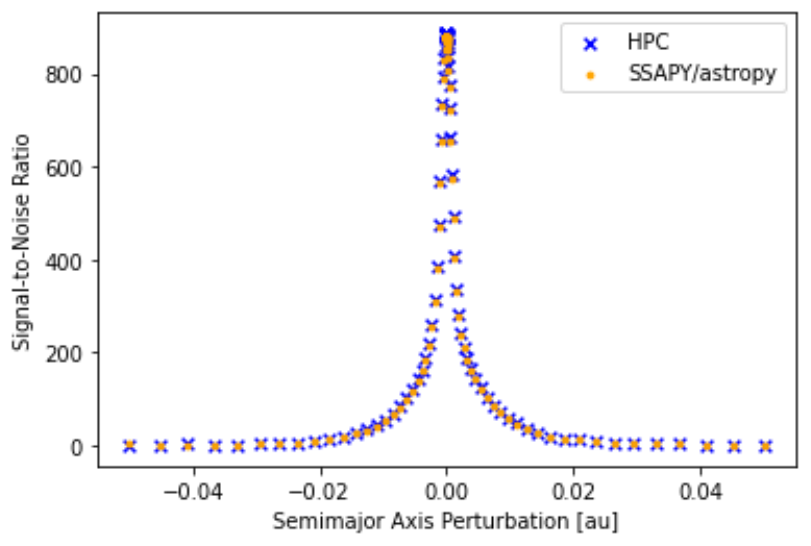}
    \caption{Signal-to-noise ratio of HPC and \texttt{SSAPy}/\texttt{astropy} asteroid detection pipelines for orbits based on a known object (Chiron) with perturbed semi-major axis values.}
    \label{fig:chiron_snr}
\end{figure}

For detection verification, we compare with the Python implementation for detecting the Centaur (2060) Chiron. In this setting, we perturb the semi-major axis of Chiron's orbit as specified by the Horizons system\footnote{\url{https://ssd.jpl.nasa.gov/horizons}} to compare in the cases of true detections and misses. Figure~\ref{fig:chiron_snr} shows the signal-to-noise ratio of both implementations in this setting, exhibiting strong agreement between the two. Interested readers can search for Chiron and follow our algorithm description in \S\ref{sec:framework} by using the ZTF forced photometry service~\citep{2023arXiv230516279M}.

\subsection{Performance Studies}
\label{sec:performance}

For applications of our non-linear digital tracking framework, we are often limited by the amount of imagery data that can be held at one time. Using four bytes per pixel, an image from ZTF takes 36~MB of space. For each terabyte of local SSD space available for storing images, we can store just over 29,000 images, much fewer than the tens of millions of image files that comprise the ZTF survey. Weak scaling studies in which each compute node is given the most data possible are the most relevant given the combination of the data volume and our intended use case. In this setting, it is worth noting that for a given set of orbits to search, doubling the number of images being tested against also roughly doubles the amount of computation required, as each orbit is intersected with all available images.

For these performance studies, we loaded 20,000 images on the SSDs attached to each compute node and tested 1.3 million orbits for detections within the available images. We focus on two separate test cases. The first is a diffuse search where a broad swath of images and orbits are used. This case is representative of a blind search for unknown asteroids. The second is a focused search in which a smaller class of orbits with a large semi-major axis is tested against a set of images chosen to be located in the direction of this collection of orbits. The parameters chosen for this focused search are based on the set of parameters identified as candidate orbits of the hypothesized planet Nine~\citep{batygin2019planet9,brown2021planet9_orbit,brown2022planet9_ztf}.

\begin{figure}[h!]
	\centering
    \includegraphics[width=\columnwidth]{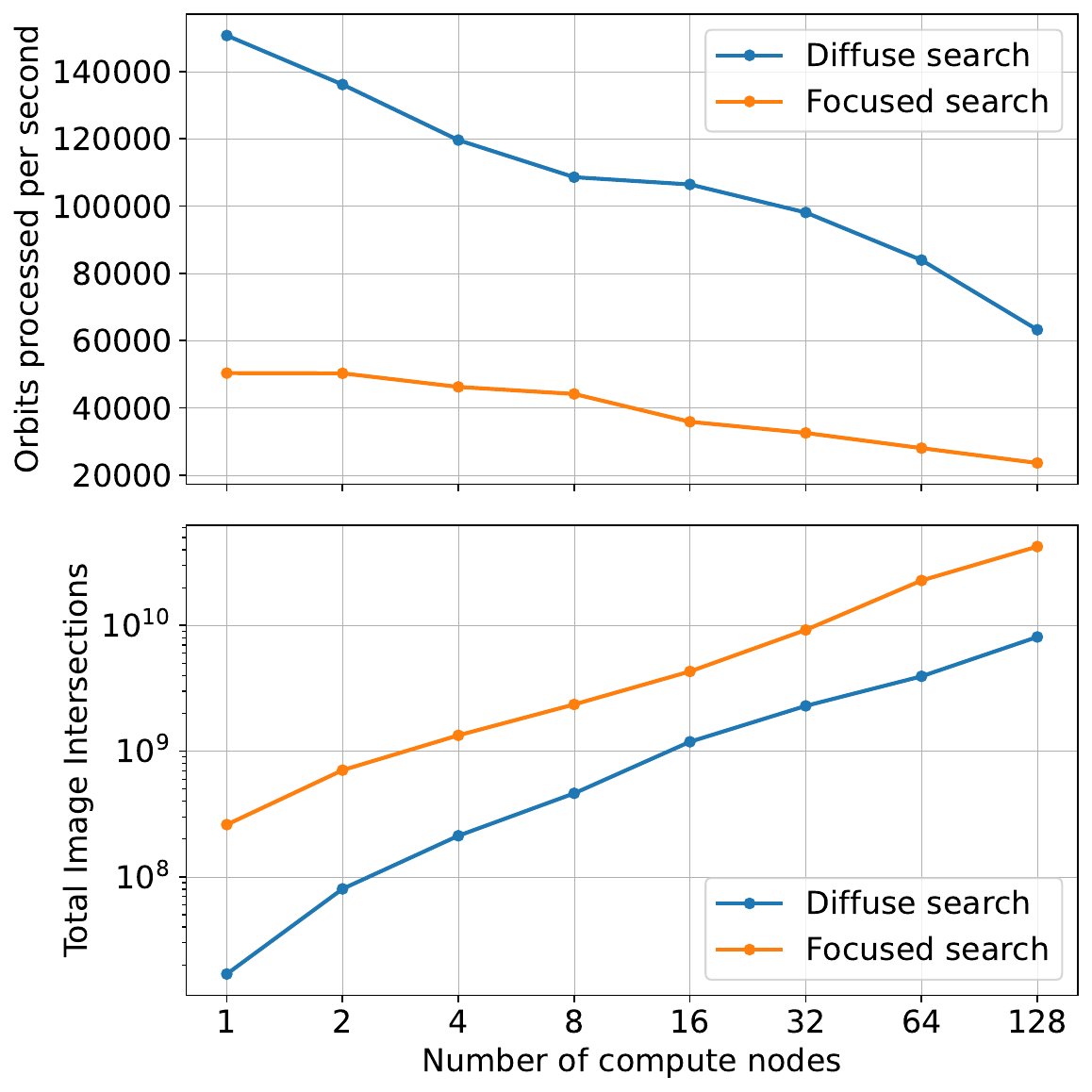}
	\caption{Weak scaling results for asteroid detection pipeline for focused and diffuse search cases in terms of the total throughput of final determinations of whether an orbit contains an asteroid (top) and number of intersections processed (bottom) in each setting.} 
	\label{fig:scaling_results}
\end{figure}

The results of our weak-scaling study in which each compute node has the same number of images assigned regardless of the number of compute nodes are given in Figure~\ref{fig:scaling_results}. In this setting, each GPU in the system is performing the same amount of work as the number of nodes increases because we are using a fixed number of orbits and the number of images per compute node remains constant. In the bottom of Figure~\ref{fig:scaling_results}, note that the number of intersections between candidate orbits and the set of images scales linearly with the number of compute nodes. This quantity reflects the amount of intersection data that must be transferred from the GPU to the host, as well as the number of images that are required to be read from each SSD. As this scales linearly with the number of compute nodes, the amount of work being performed by each compute node remains constant on average as the scale of the problem is increased.

\begin{figure}[h!]
    \centering
    \includegraphics[width=\columnwidth]{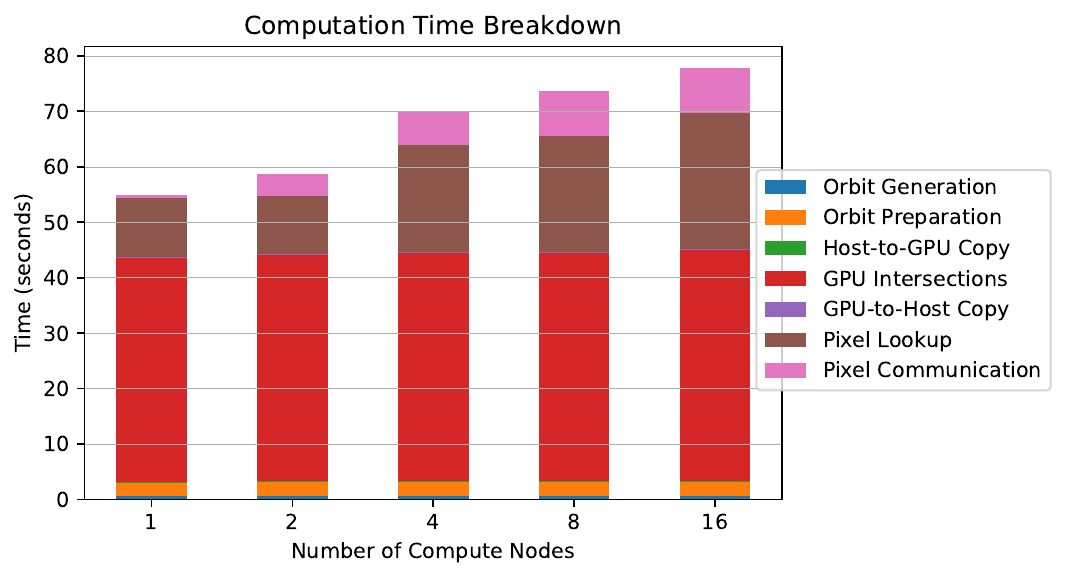}
    \caption{Timing results of individual stages of computation in a diffuse search setting. The following components are timed: Orbit Generation - time to randomly sample orbits, Orbit Preparation - time to precompute values used in all intersection calculations for a given orbit, Host-to-GPU Copy - time to copy orbital parameters from the host to the GPU, GPU Intersections - time to find positions of intersections of each orbit with every image held locally and filter out non-intersections, GPU-to-Host Copy - time to copy intersections from GPU to the host, Pixel Lookup - time to extract the necessary pixel values within images for all valid intersections, and Pixel Communication - time to communicate locally-found intersections and make a determination of whether a given orbit likely contains an object.}
    \label{fig:scaling_breakdown}
\end{figure}

Despite this fixed amount of work that each compute node performs, the top of Figure~\ref{fig:scaling_results} shows that the system throughput slowly degrades with increasing numbers of compute nodes. To investigate the scalability, Figure~\ref{fig:scaling_breakdown} contains timing results for smaller numbers of nodes in a diffuse search setting that presents the greatest loss in performance as the number of nodes is increased. In order to obtain this breakdown of timings, we must slightly alter how our code runs, which may slightly affect the end-to-end runtimes. Most notably, our code performs the Pixel Lookup and Pixel Communication steps in a single step that retrieves image data from the SSDs and queues up the information for the local intersection to be sent when the communication runtime determines it is appropriate. This allows ranks that finish retrieving image data from the SSDs faster to begin sending their results and processing incoming results sooner. By adding timers for these steps locally, ranks that begin the Pixel Communication phase early will be waiting longer because some of their results will be coming from others ranks with delayed entries to the Pixel Communication phase. To avoid this distortion in timings, we have to add an additional synchronization point between the Pixel Lookup and Pixel Communication steps that is not normally present.

Figure~\ref{fig:scaling_breakdown} shows that the largest amount of time is spent computing intersections on the GPU. This time remains constant across different node counts. The steps that we see not scaling are getting pixel values for valid intersections and communicating those results. All other phases of the computation take nearly constant amounts of time as the number of compute nodes increases. By partitioning our image metadata as described in \S\ref{sec:gpu_memory} and filtering out intersections that do not lie within the bounds of images before transferring data from the GPU back to the host processor, we get data transfer times that are a very small portion of the overall runtime.

\subsubsection{Intersection Count Variance Among Ranks}

\begin{figure}[h!]
    \centering
    \includegraphics[width=\columnwidth]{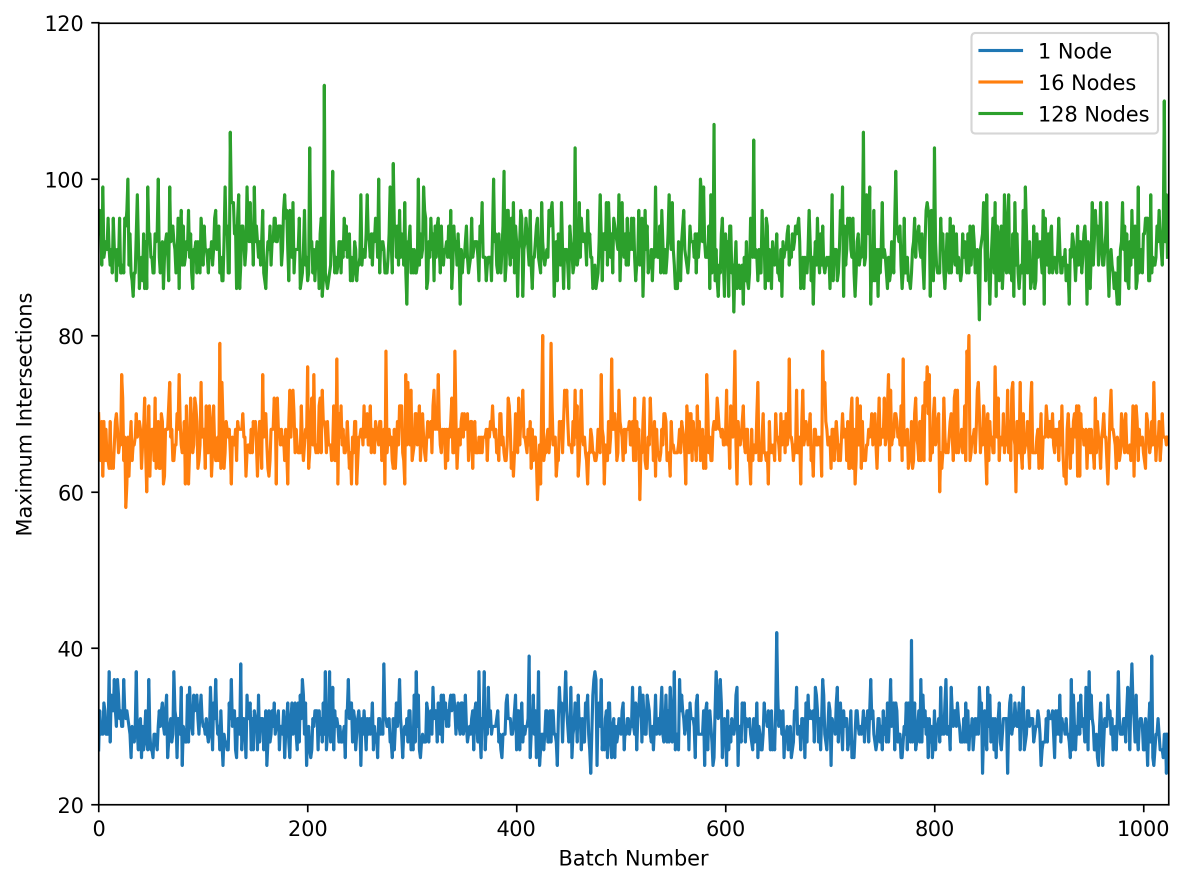}
    \caption{Maximum number of valid intersections identified on any individual MPI rank across 8000 orbits each for a blind search survey experiment. Results are shown for 1024 example batches of orbits run on different numbers of compute nodes.}
    \label{fig:batch_intersections}
\end{figure}

The increase in time required to retrieve pixel values from intersections and communicate those values is largely due to variations in the number of valid intersections that any rank finds within a given batch of orbits. Each intersection requires pixel values to be retrieved from the node-local SSDs, which is a relatively costly operation. Our performance in this phase of the calculation is bound by the slowest processor in the system. As the number of compute nodes is increased, the chance that a compute node identifies an outlier in the number of valid image intersections increases. This effect is shown Figure~\ref{fig:batch_intersections}, which shows the maximum number of valid intersections found by any rank across the 8000 orbits in a batch. On one compute node, averaged over 1024 batches of orbits, this maximum number is $\sim30$ intersections, while on 128 compute nodes it increases to $\sim91$ intersections. The effect of this on the time required to retrieve pixels is shown in Table~\ref{tab:pixel_retrieval_times}. The results shown were taken from a diffuse search experiment. The intersections were calculated on the GPUs, and the pixels were retrieved from images stored on the SSDs, but the results were not communicated to compute a final detection. The decision not to communicate results was made to isolate the amount of time that was spent waiting for the data from the SSDs and to avoid the previously described distortions to the timing results that are possible from adding additional synchronization points between the Pixel Lookup and Pixel Communication phases of Figure~\ref{fig:scaling_breakdown}. The time to retrieve pixels approximately quadrupled between the 1 compute node and 128 compute node experiments. This is primarily due to the tripling in the maximum number of valid intersections calculated in each batch when scaling across the same number of compute nodes.

\begin{table}
    \centering
    \begin{tabularx}{\columnwidth} { 
  | >{\raggedleft\arraybackslash}X 
  | >{\centering\arraybackslash}X 
  | >{\centering\arraybackslash}X 
  | >{\centering\arraybackslash}X| }
        \hline
        Compute Nodes & Intersection Time (sec) & Pixel Retrieval Time (sec) & Total Time (sec) \\
        \hline
        1 & 44.33 & 9.31 & 53.64 \\
        2 & 45.92 & 7.78 & 53.70 \\
        4 & 45.11 & 17.33 & 62.44 \\
        8 & 44.89 & 17.21 & 62.10 \\
        16 & 45.04 & 18.39 & 63.43 \\
        32 & 46.86 & 22.02 & 67.88 \\
        64 & 46.14 & 29.48 & 75.62 \\
        128 & 45.71 & 37.37 & 83.08 \\
        \hline
    \end{tabularx}
    \caption{Amount of time needed for calculating intersections between images and orbits and retrieving pixels from images (without communicating results) in a diffuse search setting.}
    \label{tab:pixel_retrieval_times}
\end{table}

In Table~\ref{tab:pixel_retrieval_times}, we show a 55\% increase in the total run time when scaling from 1 to 128 compute nodes. This is due to an increase in the time spent retrieving pixels. This corresponds to a roughly 35\% decrease in total throughput, which accounts for most of the performance degradation seen in Figure~\ref{fig:scaling_results}. This effect is only accounting for the increase in time due to the variations in the number of intersections any rank must retrieve from the local SSD. These variations will have a similar effect on the communication time because of imbalances in the amount of data individual ranks must communicate. Figure~\ref{fig:scaling_breakdown} suggests the effect on communication is the largest component of the remaining loss in scalability.

This scalability could be addressed by combining multiple batches of intersections found on the GPU in our current approach as sub-batches of much larger batches that the CPU manages. This approach will allow variations within sub-batches to average out across multiple batches and bring outliers to lower levels as we scale to more compute nodes. In our case, we cannot simply increase the batch size the GPU is processing due to memory constraints on the GPUs.

\subsection{Image Loading}

The pipeline begins with loading a large number of images on the local SSDs present on compute nodes. This step is performed once, regardless of the number of orbits about to be processed and therefore does not appear in the scalability of results in Figure~\ref{fig:scaling_results}. It is still important for the loading of images to not be prohibitively slow.

\begin{figure}[h!]
    \centering
    \includegraphics[width=\columnwidth]{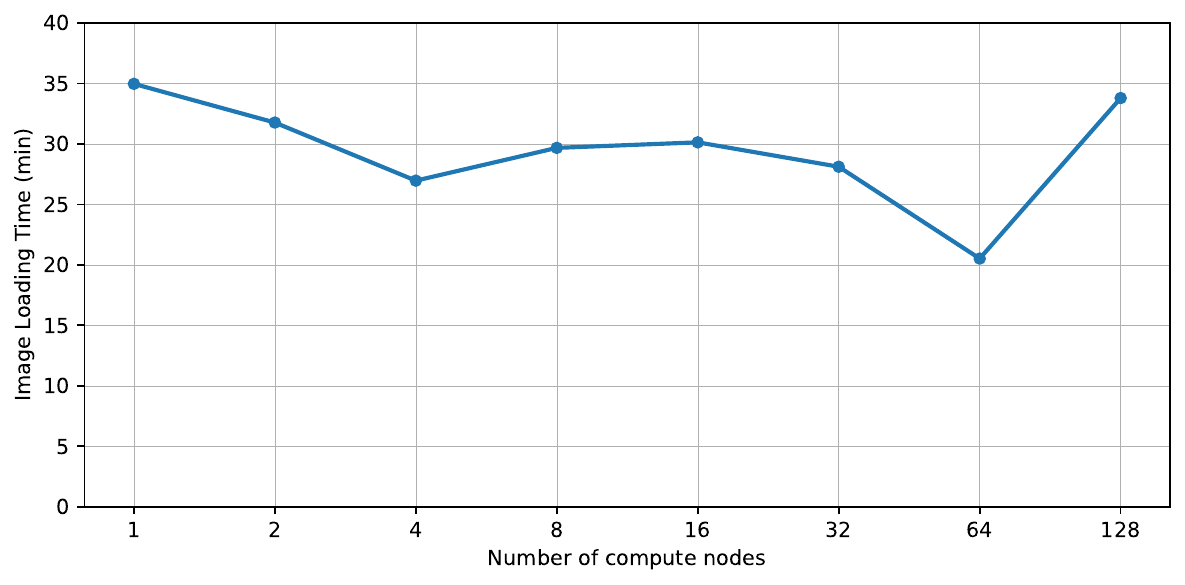}
    \caption{Time required to load 20000 images per compute node in scaling studies.}
    \label{fig:image_load_scaling}
\end{figure}

Figure~\ref{fig:image_load_scaling} shows the scaling of image loading from our experiments. We see loading times of approximately 30 minutes in length to load 20000 images onto each compute node from a parallel filesystem. These numbers are meant to give a rough idea of our expected times, but it is worth noting that this portion is very dependent on the load on the filesystem from other users.

\section{Discussion}
\label{sec:discussion}

\subsection{Summary}

We have presented a non-linear digital tracking pipeline that utilizes HPC systems to search for asteroids over timescales in which asteroid motion is highly non-linear. This system makes use of the following features available with many modern clusters:
\begin{itemize}
    \item the massive parallelism enabled by GPUs to propagate large numbers of candidate orbits and detect intersections with large imagery data sets,
    \item high-speed networks to coordinate intermediate results between processors in the system, and
    \item node-local storage on compute nodes to access very large image data sets with lower latency and higher bandwidth compared to parallel file systems.
\end{itemize}

By propagating orbits through such large data sets, we are able to combine the signal present in relatively large numbers of images. This allows for fainter detections to be made. 

We perform multiple experiments to demonstrate the accuracy and scalability of our pipeline. We compare results with a trusted implementation using \texttt{SSAPy} and \texttt{astropy} to validate intersections with individual images, as well as SNR values for a well-studied object. We have demonstrated the scalability of this approach to tackle data sets containing millions of images using up to 128 nodes on the Lassen cluster. Increasing the scale of the data set and cluster allocation is associated with a small decrease in total system throughput. 

\subsection{Future Work}

\subsubsection{Reduce Local Storage Requirements}

In our current implementation, images are partitioned across the local storage residing on each compute node. This allows us to maintain fast random access to all images simultaneously, eliminating the bottleneck of retrieving images from the parallel file system as needed.

This approach requires an allocation of compute nodes that is large enough to hold the entire data set. By partitioning the set of orbits to sample from, we could potentially reduce the portion of the data set needed at any one time, allowing a large data set to be searched with a smaller allocation of compute nodes in a more embarrassingly parallel manner. This has the benefit of more efficient access to the HPC resources when other users are present.

\subsubsection{Improve Performance and Scalability}
\label{sec:future_scalability}

As shown in Figure~\ref{fig:scaling_results}, the rate at which we can process sampled orbits slowly declines as we increase the number of compute nodes used and the number of images being searched, giving roughly 40\% of the throughput at 128 compute nodes as on a single compute node. Maintaining the same throughput while scaling the problem size and compute resources is often not a reasonable goal, but our scalability could likely be improved.

Our current implementation handles image intersections on the GPUs and the lookup of pixel values within images as distinct steps. In a pipelined implementation, we would use the GPUs to begin calculating intersections for the next batch of sampled orbits while the pixels within images are being retrieved for the current batch of orbits. Pipelining is a commonly used optimization technique that has been applied to the problems of scheduling operations within loops~\citep{allan1995pipelining}, hiding data transfer times on heterogeneous architectures to optimize matrix operations~\citep{wang2011optimizing}, and optimizing deep neural network training on multiple accelerators~\citep{huang2019gpipe}. In the context of asteroid detection, pipelining would improve performance at all scales by allowing CPUs and GPUs to work simultaneously, and it may improve the scalability by alleviating some of the effects of work imbalance within batches of orbits by removing additional synchronization points.

In addition to pipelining, decoupling the batch sizes used on the GPU and CPU can likely improve the throughput of our non-linear digital tracking framework, as discussed in Section~\ref{sec:performance}. Our current batch size is limited by the amount of memory available on GPUs, but this relatively small batch size leads to individual processes finding significantly larger numbers of valid intersections than the average process. Treating several sub-batches of GPU orbits as a single larger batch of orbits on the CPU would lower this variance and improve scalability.

Finally, another method to improve efficiency is to study the trade-offs between searching wider regions of parameter space by giving up some sensitivity. For example, images could be smoothed with a Gaussian kernel larger than the PSF, which would enable sparser sampling of orbits while still gaining signal from objects on adjacent orbits. This would enable broader coverage of orbital parameter space and more completeness at a lower sensitivity level.

\subsubsection{Interpretation of Detection Results}

The HPC non-linear digital tracking pipeline presented in this paper is able to handle data sets with millions of images and process tens of thousands of orbits per second per compute node. This is able to produce large numbers of detections, but when trialing such large numbers of orbits, noise in the data dictates that most of these are false positives caused by chance alignment positive fluctuations across images. These can be mitigated by a careful choice in threshold. More problematic are correlated noise in a few number of images caused by imperfect difference image generation. These are typically associated with bright stars. These cause outliers in the signal estimation across the intersections of an orbit.

\begin{figure}[h!]
    \centering
    \includegraphics[width=\columnwidth]{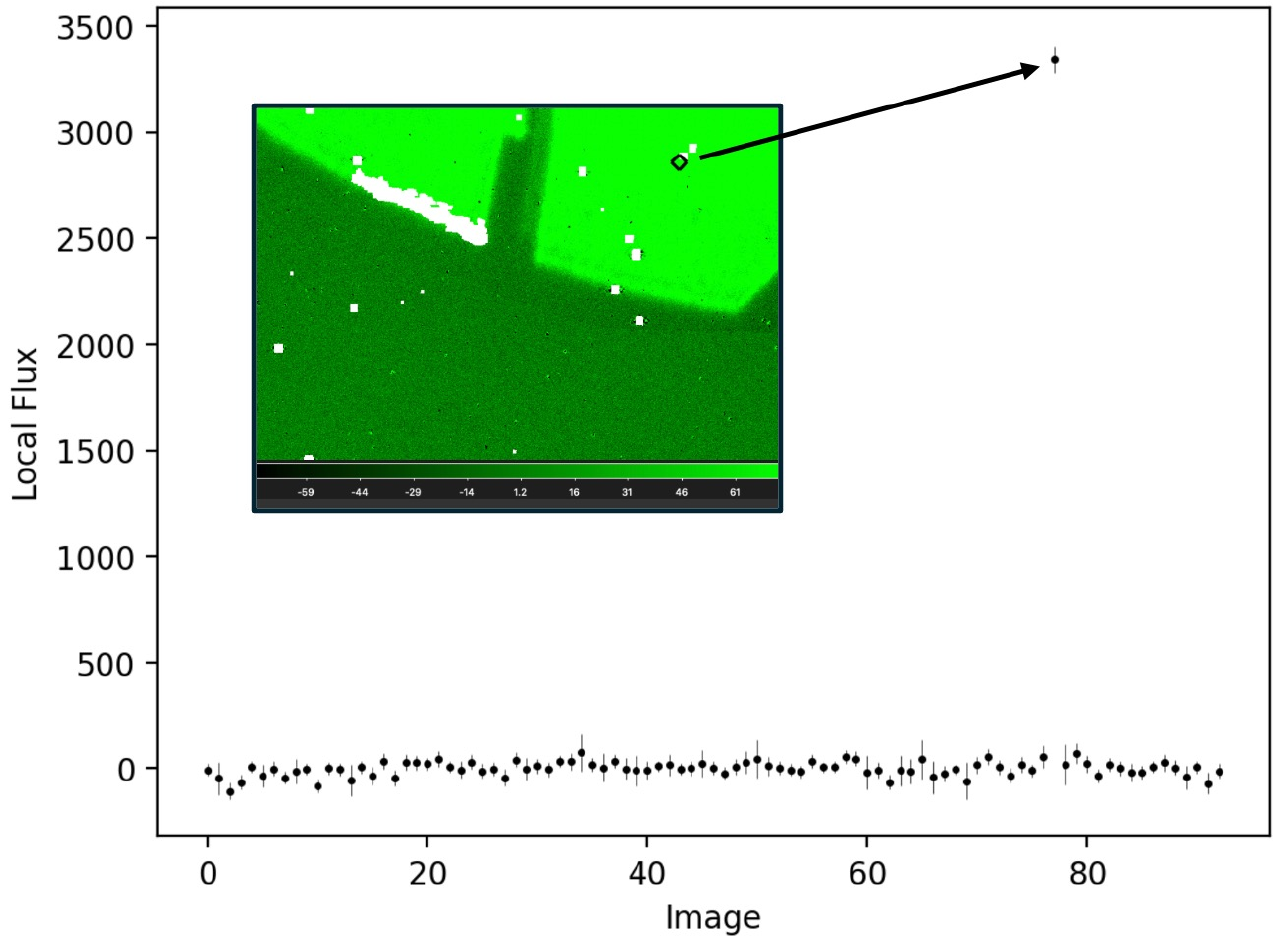}
    \caption{Example of an outlier in our pipeline caused by the chance alignment of an orbit with a difference image artifact. Simple sigma clipping removes the outlier and drops the cumulative SNR below threshold.}
    \label{fig:outlier}
\end{figure}

We are continuing to work to filter these spurious detections quickly to identify the true detections resulting from our searches. A simple masking procedure around stars from the Gaia catalog~\citep{2016A&A...595A...1G} or Pan-STARRS~\citep{2020ApJS..251....7F} appears to remove the vast majority of false positives. A more sophisticated method is to identify outliers among individual intersection SNR calculations. Intersections with anomalously high SNR value compared to the expectation can be removed. Figure~\ref{fig:outlier} shows an example, where an anomaly in the difference image of a single intersected image caused a false positive. Trimming this outlier from the total SNR calculation dropped it from 8.6 to 1.6.

Anomalies in the difference images are rare but when trialing large numbers of orbits, they do occur often enough that manipulating the images data may improve overall performance of our pipeline. Alternatively, lightcurves such as the one shown in Figure~\ref{fig:outlier} are simple to carry out sigma clipping to remove. We are also exploring filtering with supervised learning methods based on training a neural network to recognize prepared image stacks of false positives and true positives. After filtering, the remaining detections above the threshold are few, and a coordinated follow-up program with a large telescope can confirm discoveries. Regardless of which method is used, the false positive rate will ultimately need to be quantified so that detections can be contextualized. This will be studied in future work.

Another challenge in handling detections with this method is the difficulty of confirming detections without follow-up observations. We aim to use this method to detect minor planets in the outer Solar System. KBOs, for example, are important tracers of Solar System formation processes, so any detection will be worthy of dedicated follow-up observations to confirm. For larger scale surveys with our method, population studies can be carried out by injecting fake sources and estimating the recovery rate. Then, lists of detections can be used to estimate the population under the orbital distribution that was searched~\citep[e.g.][]{Bernstein04,kbo_size_distro,kbo_size_distro2}.

\subsubsection{Porting to El Capitan Systems}

El Capitan is the newest exascale Department of Energy supercomputer sited at Lawrence Livermore National Laboratory~\citep{el-capitan}. El Capitan features over 11,000 compute nodes equipped with AMD MI300A accelerated processing unit (APU) chips. These APUs are each composed of a set of CPU cores tightly coupled to three GPU chips. These CPU and GPU cores share 512 GB of high-bandwidth memory (HBM) resources, allowing direct access to memory from the CPU and GPU cores. This unified memory allows computations to proceed without expensive data transfers between CPUs and GPUs, alleviating many memory bottlenecks. In addition to computing resources, El Capitan's nodes are attached to near-node storage appliances called Rabbits~\citep{rabbits}. These Rabbits allow storage of large telescope surveys on fast local storage, much like the SSDs found on the compute nodes of Lassen.

Porting our asteroid detection pipeline to the El Capitan architecture is of great interest and has been planned for in our choice to use the RAJA performance portability library, as outlined in \S~\ref{sec:portability}. The techniques for improving scalability discussed in Section~\ref{sec:future_scalability} are specifically targeted at the Lassen architecture, centering around the limited availability of memory on discrete GPUs. The hardware on El Capitan does not face the same limitations because the GPU resources have access to each compute node's full 512 GB of HBM, a large improvement over the 64 GB of total HBM available to GPUs on each compute node of Lassen. This increased availability of memory would allow much larger batches of orbits to be tested on El Capitan systems.

Having access to the same memory space on CPUs and GPUs would allow the use of more standard producer-consumer queues in which individual compute resources operate on their data and place the results in a queue for the hardware that handles the next computational step. This would allow CPUs and GPUs to operate without explicit synchronization. Additional optimizations could be incorporated depending on the relative lengths of the work queues, such as GPUs performing additional reorganization of the valid intersections found to put them in an order more likely to yield reuse of imagery data by the CPU. Such an optimization would be especially fruitful given the larger batches of orbits that GPUs could process on El Capitan systems compared to Lassen.

These potential improvements from new APU systems could likely further improve the performance of our non-linear digital tracking framework. With its ability to process data sets that contain millions of images, we believe that this framework could be a powerful tool for minor planet detection.

\section*{Acknowledgments}
We thank the two anonymous reviewers and the editor for their helpful comments that improved the document.

This work was performed under the auspices of the U.S. Department of Energy by Lawrence Livermore National Laboratory under Contract DE-AC52-07NA27344. This work was supported by the Lawrence Livermore National Laboratory LDRD Program under Project 2023-ERD-044. The LLNL document number is LLNL-JRNL-872678.  

This work was based on observations obtained with the Samuel Oschin Telescope 48-inch and the 60-inch Telescope at the Palomar Observatory as part of the Zwicky Transient Facility project. ZTF is supported by the National Science Foundation under Grants No. AST-1440341 and AST-2034437 and a collaboration including current partners Caltech, IPAC, the Oskar Klein Center at Stockholm University, the University of Maryland, University of California, Berkeley, the University of Wisconsin at Milwaukee, University of Warwick, Ruhr University, Cornell University, Northwestern University and Drexel University. Operations are conducted by COO, IPAC, and UW.

\bibliographystyle{elsarticle-harv} 
\bibliography{bib}

\begin{thebibliography}{51}
\expandafter\ifx\csname natexlab\endcsname\relax\def\natexlab#1{#1}\fi
\providecommand{\url}[1]{\texttt{#1}}
\providecommand{\href}[2]{#2}
\providecommand{\path}[1]{#1}
\providecommand{\DOIprefix}{doi:}
\providecommand{\ArXivprefix}{arXiv:}
\providecommand{\URLprefix}{URL: }
\providecommand{\Pubmedprefix}{pmid:}
\providecommand{\doi}[1]{\href{http://dx.doi.org/#1}{\path{#1}}}
\providecommand{\Pubmed}[1]{\href{pmid:#1}{\path{#1}}}
\providecommand{\bibinfo}[2]{#2}
\ifx\xfnm\relax \def\xfnm[#1]{\unskip,\space#1}\fi
\bibitem[{Abell et~al.(2009)Abell, Allison, Anderson, Andrew, Angel, Armus, Arnett, Asztalos, Axelrod, Bailey et~al.}]{abell2009lsst}
\bibinfo{author}{Abell, P.A.}, \bibinfo{author}{Allison, J.}, \bibinfo{author}{Anderson, S.F.}, \bibinfo{author}{Andrew, J.R.}, \bibinfo{author}{Angel, J.R.P.}, \bibinfo{author}{Armus, L.}, \bibinfo{author}{Arnett, D.}, \bibinfo{author}{Asztalos, S.J.}, \bibinfo{author}{Axelrod, T.S.}, \bibinfo{author}{Bailey, S.}, et~al., \bibinfo{year}{2009}.
\newblock \bibinfo{title}{Lsst science book, version 2.0} \href{http://arxiv.org/abs/0912.0201}{{\tt arXiv:0912.0201}}.
\bibitem[{Allan et~al.(1995)Allan, Jones, Lee and Allan}]{allan1995pipelining}
\bibinfo{author}{Allan, V.H.}, \bibinfo{author}{Jones, R.B.}, \bibinfo{author}{Lee, R.M.}, \bibinfo{author}{Allan, S.J.}, \bibinfo{year}{1995}.
\newblock \bibinfo{title}{Software pipelining}.
\newblock \bibinfo{journal}{ACM Comput. Surv.} \bibinfo{volume}{27}, \bibinfo{pages}{367–432}.
\newblock \URLprefix \url{https://doi.org/10.1145/212094.212131}, \DOIprefix\doi{10.1145/212094.212131}.
\bibitem[{{Astropy Collaboration} et~al.(2022){Astropy Collaboration}, {Price-Whelan}, {Lim}, {Earl}, {Starkman}, {Bradley}, {Shupe}, {Patil}, {Corrales}, {Brasseur}, {N{"o}the}, {Donath}, {Tollerud}, {Morris}, {Ginsburg}, {Vaher}, {Weaver}, {Tocknell}, {Jamieson}, {van Kerkwijk}, {Robitaille}, {Merry}, {Bachetti}, {G{"u}nther}, {Aldcroft}, {Alvarado-Montes}, {Archibald}, {B{'o}di}, {Bapat}, {Barentsen}, {Baz{'a}n}, {Biswas}, {Boquien}, {Burke}, {Cara}, {Cara}, {Conroy}, {Conseil}, {Craig}, {Cross}, {Cruz}, {D'Eugenio}, {Dencheva}, {Devillepoix}, {Dietrich}, {Eigenbrot}, {Erben}, {Ferreira}, {Foreman-Mackey}, {Fox}, {Freij}, {Garg}, {Geda}, {Glattly}, {Gondhalekar}, {Gordon}, {Grant}, {Greenfield}, {Groener}, {Guest}, {Gurovich}, {Handberg}, {Hart}, {Hatfield-Dodds}, {Homeier}, {Hosseinzadeh}, {Jenness}, {Jones}, {Joseph}, {Kalmbach}, {Karamehmetoglu}, {Ka{l}uszy{'n}ski}, {Kelley}, {Kern}, {Kerzendorf}, {Koch}, {Kulumani}, {Lee}, {Ly}, {Ma}, {MacBride}, {Maljaars}, {Muna}, {Murphy}, {Norman}, {O'Steen},
  {Oman}, {Pacifici}, {Pascual}, {Pascual-Granado}, {Patil}, {Perren}, {Pickering}, {Rastogi}, {Roulston}, {Ryan}, {Rykoff}, {Sabater}, {Sakurikar}, {Salgado}, {Sanghi}, {Saunders}, {Savchenko}, {Schwardt}, {Seifert-Eckert}, {Shih}, {Jain}, {Shukla}, {Sick}, {Simpson}, {Singanamalla}, {Singer}, {Singhal}, {Sinha}, {Sip{H{o}}cz}, {Spitler}, {Stansby}, {Streicher}, {Sumak}, {Swinbank}, {Taranu}, {Tewary}, {Tremblay}, {Val-Borro}, {Van Kooten}, {Vasovi{'c}}, {Verma}, {de Miranda Cardoso}, {Williams}, {Wilson}, {Winkel}, {Wood-Vasey}, {Xue}, {Yoachim}, {Zhang}, {Zonca} and {Astropy Project Contributors}}]{astropy:2022}
\bibinfo{author}{{Astropy Collaboration}}, \bibinfo{author}{{Price-Whelan}, A.M.}, \bibinfo{author}{{Lim}, P.L.}, \bibinfo{author}{{Earl}, N.}, \bibinfo{author}{{Starkman}, N.}, \bibinfo{author}{{Bradley}, L.}, \bibinfo{author}{{Shupe}, D.L.}, \bibinfo{author}{{Patil}, A.A.}, \bibinfo{author}{{Corrales}, L.}, \bibinfo{author}{{Brasseur}, C.E.}, \bibinfo{author}{{N{"o}the}, M.}, \bibinfo{author}{{Donath}, A.}, \bibinfo{author}{{Tollerud}, E.}, \bibinfo{author}{{Morris}, B.M.}, \bibinfo{author}{{Ginsburg}, A.}, \bibinfo{author}{{Vaher}, E.}, \bibinfo{author}{{Weaver}, B.A.}, \bibinfo{author}{{Tocknell}, J.}, \bibinfo{author}{{Jamieson}, W.}, \bibinfo{author}{{van Kerkwijk}, M.H.}, \bibinfo{author}{{Robitaille}, T.P.}, \bibinfo{author}{{Merry}, B.}, \bibinfo{author}{{Bachetti}, M.}, \bibinfo{author}{{G{"u}nther}, H.M.}, \bibinfo{author}{{Aldcroft}, T.L.}, \bibinfo{author}{{Alvarado-Montes}, J.A.}, \bibinfo{author}{{Archibald}, A.M.}, \bibinfo{author}{{B{'o}di}, A.}, \bibinfo{author}{{Bapat}, S.},
  \bibinfo{author}{{Barentsen}, G.}, \bibinfo{author}{{Baz{'a}n}, J.}, \bibinfo{author}{{Biswas}, M.}, \bibinfo{author}{{Boquien}, M.}, \bibinfo{author}{{Burke}, D.J.}, \bibinfo{author}{{Cara}, D.}, \bibinfo{author}{{Cara}, M.}, \bibinfo{author}{{Conroy}, K.E.}, \bibinfo{author}{{Conseil}, S.}, \bibinfo{author}{{Craig}, M.W.}, \bibinfo{author}{{Cross}, R.M.}, \bibinfo{author}{{Cruz}, K.L.}, \bibinfo{author}{{D'Eugenio}, F.}, \bibinfo{author}{{Dencheva}, N.}, \bibinfo{author}{{Devillepoix}, H.A.R.}, \bibinfo{author}{{Dietrich}, J.P.}, \bibinfo{author}{{Eigenbrot}, A.D.}, \bibinfo{author}{{Erben}, T.}, \bibinfo{author}{{Ferreira}, L.}, \bibinfo{author}{{Foreman-Mackey}, D.}, \bibinfo{author}{{Fox}, R.}, \bibinfo{author}{{Freij}, N.}, \bibinfo{author}{{Garg}, S.}, \bibinfo{author}{{Geda}, R.}, \bibinfo{author}{{Glattly}, L.}, \bibinfo{author}{{Gondhalekar}, Y.}, \bibinfo{author}{{Gordon}, K.D.}, \bibinfo{author}{{Grant}, D.}, \bibinfo{author}{{Greenfield}, P.}, \bibinfo{author}{{Groener}, A.M.},
  \bibinfo{author}{{Guest}, S.}, \bibinfo{author}{{Gurovich}, S.}, \bibinfo{author}{{Handberg}, R.}, \bibinfo{author}{{Hart}, A.}, \bibinfo{author}{{Hatfield-Dodds}, Z.}, \bibinfo{author}{{Homeier}, D.}, \bibinfo{author}{{Hosseinzadeh}, G.}, \bibinfo{author}{{Jenness}, T.}, \bibinfo{author}{{Jones}, C.K.}, \bibinfo{author}{{Joseph}, P.}, \bibinfo{author}{{Kalmbach}, J.B.}, \bibinfo{author}{{Karamehmetoglu}, E.}, \bibinfo{author}{{Ka{l}uszy{'n}ski}, M.}, \bibinfo{author}{{Kelley}, M.S.P.}, \bibinfo{author}{{Kern}, N.}, \bibinfo{author}{{Kerzendorf}, W.E.}, \bibinfo{author}{{Koch}, E.W.}, \bibinfo{author}{{Kulumani}, S.}, \bibinfo{author}{{Lee}, A.}, \bibinfo{author}{{Ly}, C.}, \bibinfo{author}{{Ma}, Z.}, \bibinfo{author}{{MacBride}, C.}, \bibinfo{author}{{Maljaars}, J.M.}, \bibinfo{author}{{Muna}, D.}, \bibinfo{author}{{Murphy}, N.A.}, \bibinfo{author}{{Norman}, H.}, \bibinfo{author}{{O'Steen}, R.}, \bibinfo{author}{{Oman}, K.A.}, \bibinfo{author}{{Pacifici}, C.}, \bibinfo{author}{{Pascual}, S.},
  \bibinfo{author}{{Pascual-Granado}, J.}, \bibinfo{author}{{Patil}, R.R.}, \bibinfo{author}{{Perren}, G.I.}, \bibinfo{author}{{Pickering}, T.E.}, \bibinfo{author}{{Rastogi}, T.}, \bibinfo{author}{{Roulston}, B.R.}, \bibinfo{author}{{Ryan}, D.F.}, \bibinfo{author}{{Rykoff}, E.S.}, \bibinfo{author}{{Sabater}, J.}, \bibinfo{author}{{Sakurikar}, P.}, \bibinfo{author}{{Salgado}, J.}, \bibinfo{author}{{Sanghi}, A.}, \bibinfo{author}{{Saunders}, N.}, \bibinfo{author}{{Savchenko}, V.}, \bibinfo{author}{{Schwardt}, L.}, \bibinfo{author}{{Seifert-Eckert}, M.}, \bibinfo{author}{{Shih}, A.Y.}, \bibinfo{author}{{Jain}, A.S.}, \bibinfo{author}{{Shukla}, G.}, \bibinfo{author}{{Sick}, J.}, \bibinfo{author}{{Simpson}, C.}, \bibinfo{author}{{Singanamalla}, S.}, \bibinfo{author}{{Singer}, L.P.}, \bibinfo{author}{{Singhal}, J.}, \bibinfo{author}{{Sinha}, M.}, \bibinfo{author}{{Sip{H{o}}cz}, B.M.}, \bibinfo{author}{{Spitler}, L.R.}, \bibinfo{author}{{Stansby}, D.}, \bibinfo{author}{{Streicher}, O.}, \bibinfo{author}{{Sumak},
  J.}, \bibinfo{author}{{Swinbank}, J.D.}, \bibinfo{author}{{Taranu}, D.S.}, \bibinfo{author}{{Tewary}, N.}, \bibinfo{author}{{Tremblay}, G.R.}, \bibinfo{author}{{Val-Borro}, M.d.}, \bibinfo{author}{{Van Kooten}, S.J.}, \bibinfo{author}{{Vasovi{'c}}, Z.}, \bibinfo{author}{{Verma}, S.}, \bibinfo{author}{{de Miranda Cardoso}, J.V.}, \bibinfo{author}{{Williams}, P.K.G.}, \bibinfo{author}{{Wilson}, T.J.}, \bibinfo{author}{{Winkel}, B.}, \bibinfo{author}{{Wood-Vasey}, W.M.}, \bibinfo{author}{{Xue}, R.}, \bibinfo{author}{{Yoachim}, P.}, \bibinfo{author}{{Zhang}, C.}, \bibinfo{author}{{Zonca}, A.}, \bibinfo{author}{{Astropy Project Contributors}}, \bibinfo{year}{2022}.
\newblock \bibinfo{title}{{The Astropy Project: Sustaining and Growing a Community-oriented Open-source Project and the Latest Major Release (v5.0) of the Core Package}}.
\newblock \bibinfo{journal}{\apj} \bibinfo{volume}{935}, \bibinfo{pages}{167}.
\newblock \DOIprefix\doi{10.3847/1538-4357/ac7c74}, \href{http://arxiv.org/abs/2206.14220}{{\tt arXiv:2206.14220}}.
\bibitem[{{Astropy Collaboration} et~al.(2018){Astropy Collaboration}, {Price-Whelan}, {Sip{\H{o}}cz}, {G{\"u}nther}, {Lim}, {Crawford}, {Conseil}, {Shupe}, {Craig}, {Dencheva}, {Ginsburg}, {Vand erPlas}, {Bradley}, {P{\'e}rez-Su{\'a}rez}, {de Val-Borro}, {Aldcroft}, {Cruz}, {Robitaille}, {Tollerud}, {Ardelean}, {Babej}, {Bach}, {Bachetti}, {Bakanov}, {Bamford}, {Barentsen}, {Barmby}, {Baumbach}, {Berry}, {Biscani}, {Boquien}, {Bostroem}, {Bouma}, {Brammer}, {Bray}, {Breytenbach}, {Buddelmeijer}, {Burke}, {Calderone}, {Cano Rodr{\'\i}guez}, {Cara}, {Cardoso}, {Cheedella}, {Copin}, {Corrales}, {Crichton}, {D'Avella}, {Deil}, {Depagne}, {Dietrich}, {Donath}, {Droettboom}, {Earl}, {Erben}, {Fabbro}, {Ferreira}, {Finethy}, {Fox}, {Garrison}, {Gibbons}, {Goldstein}, {Gommers}, {Greco}, {Greenfield}, {Groener}, {Grollier}, {Hagen}, {Hirst}, {Homeier}, {Horton}, {Hosseinzadeh}, {Hu}, {Hunkeler}, {Ivezi{\'c}}, {Jain}, {Jenness}, {Kanarek}, {Kendrew}, {Kern}, {Kerzendorf}, {Khvalko}, {King}, {Kirkby}, {Kulkarni},
  {Kumar}, {Lee}, {Lenz}, {Littlefair}, {Ma}, {Macleod}, {Mastropietro}, {McCully}, {Montagnac}, {Morris}, {Mueller}, {Mumford}, {Muna}, {Murphy}, {Nelson}, {Nguyen}, {Ninan}, {N{\"o}the}, {Ogaz}, {Oh}, {Parejko}, {Parley}, {Pascual}, {Patil}, {Patil}, {Plunkett}, {Prochaska}, {Rastogi}, {Reddy Janga}, {Sabater}, {Sakurikar}, {Seifert}, {Sherbert}, {Sherwood-Taylor}, {Shih}, {Sick}, {Silbiger}, {Singanamalla}, {Singer}, {Sladen}, {Sooley}, {Sornarajah}, {Streicher}, {Teuben}, {Thomas}, {Tremblay}, {Turner}, {Terr{\'o}n}, {van Kerkwijk}, {de la Vega}, {Watkins}, {Weaver}, {Whitmore}, {Woillez}, {Zabalza} and {Astropy Contributors}}]{astropy:2018}
\bibinfo{author}{{Astropy Collaboration}}, \bibinfo{author}{{Price-Whelan}, A.M.}, \bibinfo{author}{{Sip{\H{o}}cz}, B.M.}, \bibinfo{author}{{G{\"u}nther}, H.M.}, \bibinfo{author}{{Lim}, P.L.}, \bibinfo{author}{{Crawford}, S.M.}, \bibinfo{author}{{Conseil}, S.}, \bibinfo{author}{{Shupe}, D.L.}, \bibinfo{author}{{Craig}, M.W.}, \bibinfo{author}{{Dencheva}, N.}, \bibinfo{author}{{Ginsburg}, A.}, \bibinfo{author}{{Vand erPlas}, J.T.}, \bibinfo{author}{{Bradley}, L.D.}, \bibinfo{author}{{P{\'e}rez-Su{\'a}rez}, D.}, \bibinfo{author}{{de Val-Borro}, M.}, \bibinfo{author}{{Aldcroft}, T.L.}, \bibinfo{author}{{Cruz}, K.L.}, \bibinfo{author}{{Robitaille}, T.P.}, \bibinfo{author}{{Tollerud}, E.J.}, \bibinfo{author}{{Ardelean}, C.}, \bibinfo{author}{{Babej}, T.}, \bibinfo{author}{{Bach}, Y.P.}, \bibinfo{author}{{Bachetti}, M.}, \bibinfo{author}{{Bakanov}, A.V.}, \bibinfo{author}{{Bamford}, S.P.}, \bibinfo{author}{{Barentsen}, G.}, \bibinfo{author}{{Barmby}, P.}, \bibinfo{author}{{Baumbach}, A.}, \bibinfo{author}{{Berry},
  K.L.}, \bibinfo{author}{{Biscani}, F.}, \bibinfo{author}{{Boquien}, M.}, \bibinfo{author}{{Bostroem}, K.A.}, \bibinfo{author}{{Bouma}, L.G.}, \bibinfo{author}{{Brammer}, G.B.}, \bibinfo{author}{{Bray}, E.M.}, \bibinfo{author}{{Breytenbach}, H.}, \bibinfo{author}{{Buddelmeijer}, H.}, \bibinfo{author}{{Burke}, D.J.}, \bibinfo{author}{{Calderone}, G.}, \bibinfo{author}{{Cano Rodr{\'\i}guez}, J.L.}, \bibinfo{author}{{Cara}, M.}, \bibinfo{author}{{Cardoso}, J.V.M.}, \bibinfo{author}{{Cheedella}, S.}, \bibinfo{author}{{Copin}, Y.}, \bibinfo{author}{{Corrales}, L.}, \bibinfo{author}{{Crichton}, D.}, \bibinfo{author}{{D'Avella}, D.}, \bibinfo{author}{{Deil}, C.}, \bibinfo{author}{{Depagne}, {\'E}.}, \bibinfo{author}{{Dietrich}, J.P.}, \bibinfo{author}{{Donath}, A.}, \bibinfo{author}{{Droettboom}, M.}, \bibinfo{author}{{Earl}, N.}, \bibinfo{author}{{Erben}, T.}, \bibinfo{author}{{Fabbro}, S.}, \bibinfo{author}{{Ferreira}, L.A.}, \bibinfo{author}{{Finethy}, T.}, \bibinfo{author}{{Fox}, R.T.},
  \bibinfo{author}{{Garrison}, L.H.}, \bibinfo{author}{{Gibbons}, S.L.J.}, \bibinfo{author}{{Goldstein}, D.A.}, \bibinfo{author}{{Gommers}, R.}, \bibinfo{author}{{Greco}, J.P.}, \bibinfo{author}{{Greenfield}, P.}, \bibinfo{author}{{Groener}, A.M.}, \bibinfo{author}{{Grollier}, F.}, \bibinfo{author}{{Hagen}, A.}, \bibinfo{author}{{Hirst}, P.}, \bibinfo{author}{{Homeier}, D.}, \bibinfo{author}{{Horton}, A.J.}, \bibinfo{author}{{Hosseinzadeh}, G.}, \bibinfo{author}{{Hu}, L.}, \bibinfo{author}{{Hunkeler}, J.S.}, \bibinfo{author}{{Ivezi{\'c}}, {\v{Z}}.}, \bibinfo{author}{{Jain}, A.}, \bibinfo{author}{{Jenness}, T.}, \bibinfo{author}{{Kanarek}, G.}, \bibinfo{author}{{Kendrew}, S.}, \bibinfo{author}{{Kern}, N.S.}, \bibinfo{author}{{Kerzendorf}, W.E.}, \bibinfo{author}{{Khvalko}, A.}, \bibinfo{author}{{King}, J.}, \bibinfo{author}{{Kirkby}, D.}, \bibinfo{author}{{Kulkarni}, A.M.}, \bibinfo{author}{{Kumar}, A.}, \bibinfo{author}{{Lee}, A.}, \bibinfo{author}{{Lenz}, D.}, \bibinfo{author}{{Littlefair}, S.P.},
  \bibinfo{author}{{Ma}, Z.}, \bibinfo{author}{{Macleod}, D.M.}, \bibinfo{author}{{Mastropietro}, M.}, \bibinfo{author}{{McCully}, C.}, \bibinfo{author}{{Montagnac}, S.}, \bibinfo{author}{{Morris}, B.M.}, \bibinfo{author}{{Mueller}, M.}, \bibinfo{author}{{Mumford}, S.J.}, \bibinfo{author}{{Muna}, D.}, \bibinfo{author}{{Murphy}, N.A.}, \bibinfo{author}{{Nelson}, S.}, \bibinfo{author}{{Nguyen}, G.H.}, \bibinfo{author}{{Ninan}, J.P.}, \bibinfo{author}{{N{\"o}the}, M.}, \bibinfo{author}{{Ogaz}, S.}, \bibinfo{author}{{Oh}, S.}, \bibinfo{author}{{Parejko}, J.K.}, \bibinfo{author}{{Parley}, N.}, \bibinfo{author}{{Pascual}, S.}, \bibinfo{author}{{Patil}, R.}, \bibinfo{author}{{Patil}, A.A.}, \bibinfo{author}{{Plunkett}, A.L.}, \bibinfo{author}{{Prochaska}, J.X.}, \bibinfo{author}{{Rastogi}, T.}, \bibinfo{author}{{Reddy Janga}, V.}, \bibinfo{author}{{Sabater}, J.}, \bibinfo{author}{{Sakurikar}, P.}, \bibinfo{author}{{Seifert}, M.}, \bibinfo{author}{{Sherbert}, L.E.}, \bibinfo{author}{{Sherwood-Taylor}, H.},
  \bibinfo{author}{{Shih}, A.Y.}, \bibinfo{author}{{Sick}, J.}, \bibinfo{author}{{Silbiger}, M.T.}, \bibinfo{author}{{Singanamalla}, S.}, \bibinfo{author}{{Singer}, L.P.}, \bibinfo{author}{{Sladen}, P.H.}, \bibinfo{author}{{Sooley}, K.A.}, \bibinfo{author}{{Sornarajah}, S.}, \bibinfo{author}{{Streicher}, O.}, \bibinfo{author}{{Teuben}, P.}, \bibinfo{author}{{Thomas}, S.W.}, \bibinfo{author}{{Tremblay}, G.R.}, \bibinfo{author}{{Turner}, J.E.H.}, \bibinfo{author}{{Terr{\'o}n}, V.}, \bibinfo{author}{{van Kerkwijk}, M.H.}, \bibinfo{author}{{de la Vega}, A.}, \bibinfo{author}{{Watkins}, L.L.}, \bibinfo{author}{{Weaver}, B.A.}, \bibinfo{author}{{Whitmore}, J.B.}, \bibinfo{author}{{Woillez}, J.}, \bibinfo{author}{{Zabalza}, V.}, \bibinfo{author}{{Astropy Contributors}}, \bibinfo{year}{2018}.
\newblock \bibinfo{title}{{The Astropy Project: Building an Open-science Project and Status of the v2.0 Core Package}}.
\newblock \bibinfo{journal}{\aj} \bibinfo{volume}{156}, \bibinfo{pages}{123}.
\newblock \DOIprefix\doi{10.3847/1538-3881/aabc4f}, \href{http://arxiv.org/abs/1801.02634}{{\tt arXiv:1801.02634}}.
\bibitem[{{Astropy Collaboration} et~al.(2013){Astropy Collaboration}, {Robitaille}, {Tollerud}, {Greenfield}, {Droettboom}, {Bray}, {Aldcroft}, {Davis}, {Ginsburg}, {Price-Whelan}, {Kerzendorf}, {Conley}, {Crighton}, {Barbary}, {Muna}, {Ferguson}, {Grollier}, {Parikh}, {Nair}, {Unther}, {Deil}, {Woillez}, {Conseil}, {Kramer}, {Turner}, {Singer}, {Fox}, {Weaver}, {Zabalza}, {Edwards}, {Azalee Bostroem}, {Burke}, {Casey}, {Crawford}, {Dencheva}, {Ely}, {Jenness}, {Labrie}, {Lim}, {Pierfederici}, {Pontzen}, {Ptak}, {Refsdal}, {Servillat} and {Streicher}}]{astropy:2013}
\bibinfo{author}{{Astropy Collaboration}}, \bibinfo{author}{{Robitaille}, T.P.}, \bibinfo{author}{{Tollerud}, E.J.}, \bibinfo{author}{{Greenfield}, P.}, \bibinfo{author}{{Droettboom}, M.}, \bibinfo{author}{{Bray}, E.}, \bibinfo{author}{{Aldcroft}, T.}, \bibinfo{author}{{Davis}, M.}, \bibinfo{author}{{Ginsburg}, A.}, \bibinfo{author}{{Price-Whelan}, A.M.}, \bibinfo{author}{{Kerzendorf}, W.E.}, \bibinfo{author}{{Conley}, A.}, \bibinfo{author}{{Crighton}, N.}, \bibinfo{author}{{Barbary}, K.}, \bibinfo{author}{{Muna}, D.}, \bibinfo{author}{{Ferguson}, H.}, \bibinfo{author}{{Grollier}, F.}, \bibinfo{author}{{Parikh}, M.M.}, \bibinfo{author}{{Nair}, P.H.}, \bibinfo{author}{{Unther}, H.M.}, \bibinfo{author}{{Deil}, C.}, \bibinfo{author}{{Woillez}, J.}, \bibinfo{author}{{Conseil}, S.}, \bibinfo{author}{{Kramer}, R.}, \bibinfo{author}{{Turner}, J.E.H.}, \bibinfo{author}{{Singer}, L.}, \bibinfo{author}{{Fox}, R.}, \bibinfo{author}{{Weaver}, B.A.}, \bibinfo{author}{{Zabalza}, V.}, \bibinfo{author}{{Edwards}, Z.I.},
  \bibinfo{author}{{Azalee Bostroem}, K.}, \bibinfo{author}{{Burke}, D.J.}, \bibinfo{author}{{Casey}, A.R.}, \bibinfo{author}{{Crawford}, S.M.}, \bibinfo{author}{{Dencheva}, N.}, \bibinfo{author}{{Ely}, J.}, \bibinfo{author}{{Jenness}, T.}, \bibinfo{author}{{Labrie}, K.}, \bibinfo{author}{{Lim}, P.L.}, \bibinfo{author}{{Pierfederici}, F.}, \bibinfo{author}{{Pontzen}, A.}, \bibinfo{author}{{Ptak}, A.}, \bibinfo{author}{{Refsdal}, B.}, \bibinfo{author}{{Servillat}, M.}, \bibinfo{author}{{Streicher}, O.}, \bibinfo{year}{2013}.
\newblock \bibinfo{title}{{Astropy: A community Python package for astronomy}}.
\newblock \bibinfo{journal}{\aap} \bibinfo{volume}{558}, \bibinfo{pages}{A33}.
\newblock \DOIprefix\doi{10.1051/0004-6361/201322068}, \href{http://arxiv.org/abs/1307.6212}{{\tt arXiv:1307.6212}}.
\bibitem[{Auten and Epperly(2023)}]{rabbits}
\bibinfo{author}{Auten, H.}, \bibinfo{author}{Epperly, M.}, \bibinfo{year}{2023}.
\newblock \bibinfo{title}{Road to {E}l {C}apitan 4: Storage in the exascale era}.
\newblock \bibinfo{howpublished}{\url{https://computing.llnl.gov/about/newsroom/road-el-capitan-4}}.
\bibitem[{Batygin et~al.(2019)Batygin, Adams, Brown and Becker}]{batygin2019planet9}
\bibinfo{author}{Batygin, K.}, \bibinfo{author}{Adams, F.C.}, \bibinfo{author}{Brown, M.E.}, \bibinfo{author}{Becker, J.C.}, \bibinfo{year}{2019}.
\newblock \bibinfo{title}{The planet nine hypothesis}.
\newblock \bibinfo{journal}{Physics Reports} \bibinfo{volume}{805}, \bibinfo{pages}{1--53}.
\bibitem[{Beckingsale et~al.(2019)Beckingsale, Burmark, Hornung, Jones, Killian, Kunen, Pearce, Robinson, Ryujin and Scogland}]{beckingsale2019raja}
\bibinfo{author}{Beckingsale, D.A.}, \bibinfo{author}{Burmark, J.}, \bibinfo{author}{Hornung, R.}, \bibinfo{author}{Jones, H.}, \bibinfo{author}{Killian, W.}, \bibinfo{author}{Kunen, A.J.}, \bibinfo{author}{Pearce, O.}, \bibinfo{author}{Robinson, P.}, \bibinfo{author}{Ryujin, B.S.}, \bibinfo{author}{Scogland, T.R.}, \bibinfo{year}{2019}.
\newblock \bibinfo{title}{Raja: Portable performance for large-scale scientific applications}, in: \bibinfo{booktitle}{2019 ieee/acm international workshop on performance, portability and productivity in hpc (p3hpc)}, \bibinfo{organization}{IEEE}. pp. \bibinfo{pages}{71--81}.
\bibitem[{{Bellm} et~al.(2019){Bellm}, {Kulkarni}, {Graham}, {Dekany}, {Smith}, {Riddle}, {Masci}, {Helou}, {Prince}, {Adams}, {Barbarino}, {Barlow}, {Bauer}, {Beck}, {Belicki}, {Biswas}, {Blagorodnova}, {Bodewits}, {Bolin}, {Brinnel}, {Brooke}, {Bue}, {Bulla}, {Burruss}, {Cenko}, {Chang}, {Connolly}, {Coughlin}, {Cromer}, {Cunningham}, {De}, {Delacroix}, {Desai}, {Duev}, {Eadie}, {Farnham}, {Feeney}, {Feindt}, {Flynn}, {Franckowiak}, {Frederick}, {Fremling}, {Gal-Yam}, {Gezari}, {Giomi}, {Goldstein}, {Golkhou}, {Goobar}, {Groom}, {Hacopians}, {Hale}, {Henning}, {Ho}, {Hover}, {Howell}, {Hung}, {Huppenkothen}, {Imel}, {Ip}, {Ivezi{\'c}}, {Jackson}, {Jones}, {Juric}, {Kasliwal}, {Kaspi}, {Kaye}, {Kelley}, {Kowalski}, {Kramer}, {Kupfer}, {Landry}, {Laher}, {Lee}, {Lin}, {Lin}, {Lunnan}, {Giomi}, {Mahabal}, {Mao}, {Miller}, {Monkewitz}, {Murphy}, {Ngeow}, {Nordin}, {Nugent}, {Ofek}, {Patterson}, {Penprase}, {Porter}, {Rauch}, {Rebbapragada}, {Reiley}, {Rigault}, {Rodriguez}, {van Roestel}, {Rusholme}, {van
  Santen}, {Schulze}, {Shupe}, {Singer}, {Soumagnac}, {Stein}, {Surace}, {Sollerman}, {Szkody}, {Taddia}, {Terek}, {Van Sistine}, {van Velzen}, {Vestrand}, {Walters}, {Ward}, {Ye}, {Yu}, {Yan} and {Zolkower}}]{ztf}
\bibinfo{author}{{Bellm}, E.C.}, \bibinfo{author}{{Kulkarni}, S.R.}, \bibinfo{author}{{Graham}, M.J.}, \bibinfo{author}{{Dekany}, R.}, \bibinfo{author}{{Smith}, R.M.}, \bibinfo{author}{{Riddle}, R.}, \bibinfo{author}{{Masci}, F.J.}, \bibinfo{author}{{Helou}, G.}, \bibinfo{author}{{Prince}, T.A.}, \bibinfo{author}{{Adams}, S.M.}, \bibinfo{author}{{Barbarino}, C.}, \bibinfo{author}{{Barlow}, T.}, \bibinfo{author}{{Bauer}, J.}, \bibinfo{author}{{Beck}, R.}, \bibinfo{author}{{Belicki}, J.}, \bibinfo{author}{{Biswas}, R.}, \bibinfo{author}{{Blagorodnova}, N.}, \bibinfo{author}{{Bodewits}, D.}, \bibinfo{author}{{Bolin}, B.}, \bibinfo{author}{{Brinnel}, V.}, \bibinfo{author}{{Brooke}, T.}, \bibinfo{author}{{Bue}, B.}, \bibinfo{author}{{Bulla}, M.}, \bibinfo{author}{{Burruss}, R.}, \bibinfo{author}{{Cenko}, S.B.}, \bibinfo{author}{{Chang}, C.K.}, \bibinfo{author}{{Connolly}, A.}, \bibinfo{author}{{Coughlin}, M.}, \bibinfo{author}{{Cromer}, J.}, \bibinfo{author}{{Cunningham}, V.}, \bibinfo{author}{{De}, K.},
  \bibinfo{author}{{Delacroix}, A.}, \bibinfo{author}{{Desai}, V.}, \bibinfo{author}{{Duev}, D.A.}, \bibinfo{author}{{Eadie}, G.}, \bibinfo{author}{{Farnham}, T.L.}, \bibinfo{author}{{Feeney}, M.}, \bibinfo{author}{{Feindt}, U.}, \bibinfo{author}{{Flynn}, D.}, \bibinfo{author}{{Franckowiak}, A.}, \bibinfo{author}{{Frederick}, S.}, \bibinfo{author}{{Fremling}, C.}, \bibinfo{author}{{Gal-Yam}, A.}, \bibinfo{author}{{Gezari}, S.}, \bibinfo{author}{{Giomi}, M.}, \bibinfo{author}{{Goldstein}, D.A.}, \bibinfo{author}{{Golkhou}, V.Z.}, \bibinfo{author}{{Goobar}, A.}, \bibinfo{author}{{Groom}, S.}, \bibinfo{author}{{Hacopians}, E.}, \bibinfo{author}{{Hale}, D.}, \bibinfo{author}{{Henning}, J.}, \bibinfo{author}{{Ho}, A.Y.Q.}, \bibinfo{author}{{Hover}, D.}, \bibinfo{author}{{Howell}, J.}, \bibinfo{author}{{Hung}, T.}, \bibinfo{author}{{Huppenkothen}, D.}, \bibinfo{author}{{Imel}, D.}, \bibinfo{author}{{Ip}, W.H.}, \bibinfo{author}{{Ivezi{\'c}}, {\v{Z}}.}, \bibinfo{author}{{Jackson}, E.}, \bibinfo{author}{{Jones}, L.},
  \bibinfo{author}{{Juric}, M.}, \bibinfo{author}{{Kasliwal}, M.M.}, \bibinfo{author}{{Kaspi}, S.}, \bibinfo{author}{{Kaye}, S.}, \bibinfo{author}{{Kelley}, M.S.P.}, \bibinfo{author}{{Kowalski}, M.}, \bibinfo{author}{{Kramer}, E.}, \bibinfo{author}{{Kupfer}, T.}, \bibinfo{author}{{Landry}, W.}, \bibinfo{author}{{Laher}, R.R.}, \bibinfo{author}{{Lee}, C.D.}, \bibinfo{author}{{Lin}, H.W.}, \bibinfo{author}{{Lin}, Z.Y.}, \bibinfo{author}{{Lunnan}, R.}, \bibinfo{author}{{Giomi}, M.}, \bibinfo{author}{{Mahabal}, A.}, \bibinfo{author}{{Mao}, P.}, \bibinfo{author}{{Miller}, A.A.}, \bibinfo{author}{{Monkewitz}, S.}, \bibinfo{author}{{Murphy}, P.}, \bibinfo{author}{{Ngeow}, C.C.}, \bibinfo{author}{{Nordin}, J.}, \bibinfo{author}{{Nugent}, P.}, \bibinfo{author}{{Ofek}, E.}, \bibinfo{author}{{Patterson}, M.T.}, \bibinfo{author}{{Penprase}, B.}, \bibinfo{author}{{Porter}, M.}, \bibinfo{author}{{Rauch}, L.}, \bibinfo{author}{{Rebbapragada}, U.}, \bibinfo{author}{{Reiley}, D.}, \bibinfo{author}{{Rigault}, M.},
  \bibinfo{author}{{Rodriguez}, H.}, \bibinfo{author}{{van Roestel}, J.}, \bibinfo{author}{{Rusholme}, B.}, \bibinfo{author}{{van Santen}, J.}, \bibinfo{author}{{Schulze}, S.}, \bibinfo{author}{{Shupe}, D.L.}, \bibinfo{author}{{Singer}, L.P.}, \bibinfo{author}{{Soumagnac}, M.T.}, \bibinfo{author}{{Stein}, R.}, \bibinfo{author}{{Surace}, J.}, \bibinfo{author}{{Sollerman}, J.}, \bibinfo{author}{{Szkody}, P.}, \bibinfo{author}{{Taddia}, F.}, \bibinfo{author}{{Terek}, S.}, \bibinfo{author}{{Van Sistine}, A.}, \bibinfo{author}{{van Velzen}, S.}, \bibinfo{author}{{Vestrand}, W.T.}, \bibinfo{author}{{Walters}, R.}, \bibinfo{author}{{Ward}, C.}, \bibinfo{author}{{Ye}, Q.Z.}, \bibinfo{author}{{Yu}, P.C.}, \bibinfo{author}{{Yan}, L.}, \bibinfo{author}{{Zolkower}, J.}, \bibinfo{year}{2019}.
\newblock \bibinfo{title}{{The Zwicky Transient Facility: System Overview, Performance, and First Results}}.
\newblock \bibinfo{journal}{\pasp} \bibinfo{volume}{131}, \bibinfo{pages}{018002}.
\newblock \DOIprefix\doi{10.1088/1538-3873/aaecbe}, \href{http://arxiv.org/abs/1902.01932}{{\tt arXiv:1902.01932}}.
\bibitem[{{Bernstein} et~al.(2004){Bernstein}, {Trilling}, {Allen}, {Brown}, {Holman} and {Malhotra}}]{Bernstein04}
\bibinfo{author}{{Bernstein}, G.M.}, \bibinfo{author}{{Trilling}, D.E.}, \bibinfo{author}{{Allen}, R.L.}, \bibinfo{author}{{Brown}, M.E.}, \bibinfo{author}{{Holman}, M.}, \bibinfo{author}{{Malhotra}, R.}, \bibinfo{year}{2004}.
\newblock \bibinfo{title}{{The Size Distribution of Trans-Neptunian Bodies}}.
\newblock \bibinfo{journal}{\aj} \bibinfo{volume}{128}, \bibinfo{pages}{1364--1390}.
\newblock \DOIprefix\doi{10.1086/422919}, \href{http://arxiv.org/abs/astro-ph/0308467}{{\tt arXiv:astro-ph/0308467}}.
\bibitem[{{Bowell} et~al.(1989){Bowell}, {Hapke}, {Domingue}, {Lumme}, {Peltoniemi} and {Harris}}]{1989aste.conf..524B}
\bibinfo{author}{{Bowell}, E.}, \bibinfo{author}{{Hapke}, B.}, \bibinfo{author}{{Domingue}, D.}, \bibinfo{author}{{Lumme}, K.}, \bibinfo{author}{{Peltoniemi}, J.}, \bibinfo{author}{{Harris}, A.W.}, \bibinfo{year}{1989}.
\newblock \bibinfo{title}{{Application of photometric models to asteroids.}}, in: \bibinfo{editor}{{Binzel}, R.P.}, \bibinfo{editor}{{Gehrels}, T.}, \bibinfo{editor}{{Matthews}, M.S.} (Eds.), \bibinfo{booktitle}{Asteroids II}, pp. \bibinfo{pages}{524--556}.
\bibitem[{Brown and Batygin(2021)}]{brown2021planet9_orbit}
\bibinfo{author}{Brown, M.E.}, \bibinfo{author}{Batygin, K.}, \bibinfo{year}{2021}.
\newblock \bibinfo{title}{The orbit of planet nine}.
\newblock \bibinfo{journal}{The Astronomical Journal} \bibinfo{volume}{162}, \bibinfo{pages}{219}.
\bibitem[{Brown and Batygin(2022)}]{brown2022planet9_ztf}
\bibinfo{author}{Brown, M.E.}, \bibinfo{author}{Batygin, K.}, \bibinfo{year}{2022}.
\newblock \bibinfo{title}{A search for planet nine using the zwicky transient facility public archive}.
\newblock \bibinfo{journal}{The Astronomical Journal} \bibinfo{volume}{163}, \bibinfo{pages}{102}.
\bibitem[{{Calabretta}(2011)}]{wcslib}
\bibinfo{author}{{Calabretta}, M.R.}, \bibinfo{year}{2011}.
\newblock \bibinfo{title}{{Wcslib and Pgsbox}}.
\newblock \bibinfo{howpublished}{Astrophysics Source Code Library, record ascl:1108.003}.
\bibitem[{{Calabretta} and {Greisen}(2002)}]{2002A&A...395.1077C}
\bibinfo{author}{{Calabretta}, M.R.}, \bibinfo{author}{{Greisen}, E.W.}, \bibinfo{year}{2002}.
\newblock \bibinfo{title}{{Representations of celestial coordinates in FITS}}.
\newblock \bibinfo{journal}{\aap} \bibinfo{volume}{395}, \bibinfo{pages}{1077--1122}.
\newblock \DOIprefix\doi{10.1051/0004-6361:20021327}, \href{http://arxiv.org/abs/astro-ph/0207413}{{\tt arXiv:astro-ph/0207413}}.
\bibitem[{{Carter Edwards} et~al.(2014){Carter Edwards}, Trott and Sunderland}]{edwards2014kokkos}
\bibinfo{author}{{Carter Edwards}, H.}, \bibinfo{author}{Trott, C.R.}, \bibinfo{author}{Sunderland, D.}, \bibinfo{year}{2014}.
\newblock \bibinfo{title}{Kokkos: Enabling manycore performance portability through polymorphic memory access patterns}.
\newblock \bibinfo{journal}{Journal of Parallel and Distributed Computing} \bibinfo{volume}{74}, \bibinfo{pages}{3202--3216}.
\newblock \URLprefix \url{https://www.sciencedirect.com/science/article/pii/S0743731514001257}, \DOIprefix\doi{https://doi.org/10.1016/j.jpdc.2014.07.003}. \bibinfo{note}{domain-Specific Languages and High-Level Frameworks for High-Performance Computing}.
\bibitem[{{Chambers} et~al.(2016){Chambers}, {Magnier}, {Metcalfe}, {Flewelling}, {Huber}, {Waters}, {Denneau}, {Draper}, {Farrow}, {Finkbeiner}, {Holmberg}, {Koppenhoefer}, {Price}, {Rest}, {Saglia}, {Schlafly}, {Smartt}, {Sweeney}, {Wainscoat}, {Burgett}, {Chastel}, {Grav}, {Heasley}, {Hodapp}, {Jedicke}, {Kaiser}, {Kudritzki}, {Luppino}, {Lupton}, {Monet}, {Morgan}, {Onaka}, {Shiao}, {Stubbs}, {Tonry}, {White}, {Ba{\~n}ados}, {Bell}, {Bender}, {Bernard}, {Boegner}, {Boffi}, {Botticella}, {Calamida}, {Casertano}, {Chen}, {Chen}, {Cole}, {Deacon}, {Frenk}, {Fitzsimmons}, {Gezari}, {Gibbs}, {Goessl}, {Goggia}, {Gourgue}, {Goldman}, {Grant}, {Grebel}, {Hambly}, {Hasinger}, {Heavens}, {Heckman}, {Henderson}, {Henning}, {Holman}, {Hopp}, {Ip}, {Isani}, {Jackson}, {Keyes}, {Koekemoer}, {Kotak}, {Le}, {Liska}, {Long}, {Lucey}, {Liu}, {Martin}, {Masci}, {McLean}, {Mindel}, {Misra}, {Morganson}, {Murphy}, {Obaika}, {Narayan}, {Nieto-Santisteban}, {Norberg}, {Peacock}, {Pier}, {Postman}, {Primak}, {Rae}, {Rai},
  {Riess}, {Riffeser}, {Rix}, {R{\"o}ser}, {Russel}, {Rutz}, {Schilbach}, {Schultz}, {Scolnic}, {Strolger}, {Szalay}, {Seitz}, {Small}, {Smith}, {Soderblom}, {Taylor}, {Thomson}, {Taylor}, {Thakar}, {Thiel}, {Thilker}, {Unger}, {Urata}, {Valenti}, {Wagner}, {Walder}, {Walter}, {Watters}, {Werner}, {Wood-Vasey} and {Wyse}}]{panstarrs}
\bibinfo{author}{{Chambers}, K.C.}, \bibinfo{author}{{Magnier}, E.A.}, \bibinfo{author}{{Metcalfe}, N.}, \bibinfo{author}{{Flewelling}, H.A.}, \bibinfo{author}{{Huber}, M.E.}, \bibinfo{author}{{Waters}, C.Z.}, \bibinfo{author}{{Denneau}, L.}, \bibinfo{author}{{Draper}, P.W.}, \bibinfo{author}{{Farrow}, D.}, \bibinfo{author}{{Finkbeiner}, D.P.}, \bibinfo{author}{{Holmberg}, C.}, \bibinfo{author}{{Koppenhoefer}, J.}, \bibinfo{author}{{Price}, P.A.}, \bibinfo{author}{{Rest}, A.}, \bibinfo{author}{{Saglia}, R.P.}, \bibinfo{author}{{Schlafly}, E.F.}, \bibinfo{author}{{Smartt}, S.J.}, \bibinfo{author}{{Sweeney}, W.}, \bibinfo{author}{{Wainscoat}, R.J.}, \bibinfo{author}{{Burgett}, W.S.}, \bibinfo{author}{{Chastel}, S.}, \bibinfo{author}{{Grav}, T.}, \bibinfo{author}{{Heasley}, J.N.}, \bibinfo{author}{{Hodapp}, K.W.}, \bibinfo{author}{{Jedicke}, R.}, \bibinfo{author}{{Kaiser}, N.}, \bibinfo{author}{{Kudritzki}, R.P.}, \bibinfo{author}{{Luppino}, G.A.}, \bibinfo{author}{{Lupton}, R.H.}, \bibinfo{author}{{Monet},
  D.G.}, \bibinfo{author}{{Morgan}, J.S.}, \bibinfo{author}{{Onaka}, P.M.}, \bibinfo{author}{{Shiao}, B.}, \bibinfo{author}{{Stubbs}, C.W.}, \bibinfo{author}{{Tonry}, J.L.}, \bibinfo{author}{{White}, R.}, \bibinfo{author}{{Ba{\~n}ados}, E.}, \bibinfo{author}{{Bell}, E.F.}, \bibinfo{author}{{Bender}, R.}, \bibinfo{author}{{Bernard}, E.J.}, \bibinfo{author}{{Boegner}, M.}, \bibinfo{author}{{Boffi}, F.}, \bibinfo{author}{{Botticella}, M.T.}, \bibinfo{author}{{Calamida}, A.}, \bibinfo{author}{{Casertano}, S.}, \bibinfo{author}{{Chen}, W.P.}, \bibinfo{author}{{Chen}, X.}, \bibinfo{author}{{Cole}, S.}, \bibinfo{author}{{Deacon}, N.}, \bibinfo{author}{{Frenk}, C.}, \bibinfo{author}{{Fitzsimmons}, A.}, \bibinfo{author}{{Gezari}, S.}, \bibinfo{author}{{Gibbs}, V.}, \bibinfo{author}{{Goessl}, C.}, \bibinfo{author}{{Goggia}, T.}, \bibinfo{author}{{Gourgue}, R.}, \bibinfo{author}{{Goldman}, B.}, \bibinfo{author}{{Grant}, P.}, \bibinfo{author}{{Grebel}, E.K.}, \bibinfo{author}{{Hambly}, N.C.}, \bibinfo{author}{{Hasinger},
  G.}, \bibinfo{author}{{Heavens}, A.F.}, \bibinfo{author}{{Heckman}, T.M.}, \bibinfo{author}{{Henderson}, R.}, \bibinfo{author}{{Henning}, T.}, \bibinfo{author}{{Holman}, M.}, \bibinfo{author}{{Hopp}, U.}, \bibinfo{author}{{Ip}, W.H.}, \bibinfo{author}{{Isani}, S.}, \bibinfo{author}{{Jackson}, M.}, \bibinfo{author}{{Keyes}, C.D.}, \bibinfo{author}{{Koekemoer}, A.M.}, \bibinfo{author}{{Kotak}, R.}, \bibinfo{author}{{Le}, D.}, \bibinfo{author}{{Liska}, D.}, \bibinfo{author}{{Long}, K.S.}, \bibinfo{author}{{Lucey}, J.R.}, \bibinfo{author}{{Liu}, M.}, \bibinfo{author}{{Martin}, N.F.}, \bibinfo{author}{{Masci}, G.}, \bibinfo{author}{{McLean}, B.}, \bibinfo{author}{{Mindel}, E.}, \bibinfo{author}{{Misra}, P.}, \bibinfo{author}{{Morganson}, E.}, \bibinfo{author}{{Murphy}, D.N.A.}, \bibinfo{author}{{Obaika}, A.}, \bibinfo{author}{{Narayan}, G.}, \bibinfo{author}{{Nieto-Santisteban}, M.A.}, \bibinfo{author}{{Norberg}, P.}, \bibinfo{author}{{Peacock}, J.A.}, \bibinfo{author}{{Pier}, E.A.}, \bibinfo{author}{{Postman},
  M.}, \bibinfo{author}{{Primak}, N.}, \bibinfo{author}{{Rae}, C.}, \bibinfo{author}{{Rai}, A.}, \bibinfo{author}{{Riess}, A.}, \bibinfo{author}{{Riffeser}, A.}, \bibinfo{author}{{Rix}, H.W.}, \bibinfo{author}{{R{\"o}ser}, S.}, \bibinfo{author}{{Russel}, R.}, \bibinfo{author}{{Rutz}, L.}, \bibinfo{author}{{Schilbach}, E.}, \bibinfo{author}{{Schultz}, A.S.B.}, \bibinfo{author}{{Scolnic}, D.}, \bibinfo{author}{{Strolger}, L.}, \bibinfo{author}{{Szalay}, A.}, \bibinfo{author}{{Seitz}, S.}, \bibinfo{author}{{Small}, E.}, \bibinfo{author}{{Smith}, K.W.}, \bibinfo{author}{{Soderblom}, D.R.}, \bibinfo{author}{{Taylor}, P.}, \bibinfo{author}{{Thomson}, R.}, \bibinfo{author}{{Taylor}, A.N.}, \bibinfo{author}{{Thakar}, A.R.}, \bibinfo{author}{{Thiel}, J.}, \bibinfo{author}{{Thilker}, D.}, \bibinfo{author}{{Unger}, D.}, \bibinfo{author}{{Urata}, Y.}, \bibinfo{author}{{Valenti}, J.}, \bibinfo{author}{{Wagner}, J.}, \bibinfo{author}{{Walder}, T.}, \bibinfo{author}{{Walter}, F.}, \bibinfo{author}{{Watters}, S.P.},
  \bibinfo{author}{{Werner}, S.}, \bibinfo{author}{{Wood-Vasey}, W.M.}, \bibinfo{author}{{Wyse}, R.}, \bibinfo{year}{2016}.
\newblock \bibinfo{title}{{The Pan-STARRS1 Surveys}}.
\newblock \bibinfo{journal}{arXiv e-prints} , \bibinfo{pages}{arXiv:1612.05560}\DOIprefix\doi{10.48550/arXiv.1612.05560}, \href{http://arxiv.org/abs/1612.05560}{{\tt arXiv:1612.05560}}.
\bibitem[{{Cochran} et~al.(1995){Cochran}, {Levison}, {Stern} and {Duncan}}]{Cochran95}
\bibinfo{author}{{Cochran}, A.L.}, \bibinfo{author}{{Levison}, H.F.}, \bibinfo{author}{{Stern}, S.A.}, \bibinfo{author}{{Duncan}, M.J.}, \bibinfo{year}{1995}.
\newblock \bibinfo{title}{{The Discovery of Halley-sized Kuiper Belt Objects Using the Hubble Space Telescope}}.
\newblock \bibinfo{journal}{\apj} \bibinfo{volume}{455}, \bibinfo{pages}{342}.
\newblock \DOIprefix\doi{10.1086/176581}, \href{http://arxiv.org/abs/astro-ph/9509100}{{\tt arXiv:astro-ph/9509100}}.
\bibitem[{{Drake} et~al.(2009){Drake}, {Djorgovski}, {Mahabal}, {Beshore}, {Larson}, {Graham}, {Williams}, {Christensen}, {Catelan}, {Boattini}, {Gibbs}, {Hill} and {Kowalski}}]{CSS}
\bibinfo{author}{{Drake}, A.J.}, \bibinfo{author}{{Djorgovski}, S.G.}, \bibinfo{author}{{Mahabal}, A.}, \bibinfo{author}{{Beshore}, E.}, \bibinfo{author}{{Larson}, S.}, \bibinfo{author}{{Graham}, M.J.}, \bibinfo{author}{{Williams}, R.}, \bibinfo{author}{{Christensen}, E.}, \bibinfo{author}{{Catelan}, M.}, \bibinfo{author}{{Boattini}, A.}, \bibinfo{author}{{Gibbs}, A.}, \bibinfo{author}{{Hill}, R.}, \bibinfo{author}{{Kowalski}, R.}, \bibinfo{year}{2009}.
\newblock \bibinfo{title}{{First Results from the Catalina Real-Time Transient Survey}}.
\newblock \bibinfo{journal}{\apj} \bibinfo{volume}{696}, \bibinfo{pages}{870--884}.
\newblock \DOIprefix\doi{10.1088/0004-637X/696/1/870}, \href{http://arxiv.org/abs/0809.1394}{{\tt arXiv:0809.1394}}.
\bibitem[{{Flewelling} et~al.(2020){Flewelling}, {Magnier}, {Chambers}, {Heasley}, {Holmberg}, {Huber}, {Sweeney}, {Waters}, {Calamida}, {Casertano}, {Chen}, {Farrow}, {Hasinger}, {Henderson}, {Long}, {Metcalfe}, {Narayan}, {Nieto-Santisteban}, {Norberg}, {Rest}, {Saglia}, {Szalay}, {Thakar}, {Tonry}, {Valenti}, {Werner}, {White}, {Denneau}, {Draper}, {Hodapp}, {Jedicke}, {Kaiser}, {Kudritzki}, {Price}, {Wainscoat}, {Chastel}, {McLean}, {Postman} and {Shiao}}]{2020ApJS..251....7F}
\bibinfo{author}{{Flewelling}, H.A.}, \bibinfo{author}{{Magnier}, E.A.}, \bibinfo{author}{{Chambers}, K.C.}, \bibinfo{author}{{Heasley}, J.N.}, \bibinfo{author}{{Holmberg}, C.}, \bibinfo{author}{{Huber}, M.E.}, \bibinfo{author}{{Sweeney}, W.}, \bibinfo{author}{{Waters}, C.Z.}, \bibinfo{author}{{Calamida}, A.}, \bibinfo{author}{{Casertano}, S.}, \bibinfo{author}{{Chen}, X.}, \bibinfo{author}{{Farrow}, D.}, \bibinfo{author}{{Hasinger}, G.}, \bibinfo{author}{{Henderson}, R.}, \bibinfo{author}{{Long}, K.S.}, \bibinfo{author}{{Metcalfe}, N.}, \bibinfo{author}{{Narayan}, G.}, \bibinfo{author}{{Nieto-Santisteban}, M.A.}, \bibinfo{author}{{Norberg}, P.}, \bibinfo{author}{{Rest}, A.}, \bibinfo{author}{{Saglia}, R.P.}, \bibinfo{author}{{Szalay}, A.}, \bibinfo{author}{{Thakar}, A.R.}, \bibinfo{author}{{Tonry}, J.L.}, \bibinfo{author}{{Valenti}, J.}, \bibinfo{author}{{Werner}, S.}, \bibinfo{author}{{White}, R.}, \bibinfo{author}{{Denneau}, L.}, \bibinfo{author}{{Draper}, P.W.}, \bibinfo{author}{{Hodapp}, K.W.},
  \bibinfo{author}{{Jedicke}, R.}, \bibinfo{author}{{Kaiser}, N.}, \bibinfo{author}{{Kudritzki}, R.P.}, \bibinfo{author}{{Price}, P.A.}, \bibinfo{author}{{Wainscoat}, R.J.}, \bibinfo{author}{{Chastel}, S.}, \bibinfo{author}{{McLean}, B.}, \bibinfo{author}{{Postman}, M.}, \bibinfo{author}{{Shiao}, B.}, \bibinfo{year}{2020}.
\newblock \bibinfo{title}{{The Pan-STARRS1 Database and Data Products}}.
\newblock \bibinfo{journal}{\apjs} \bibinfo{volume}{251}, \bibinfo{pages}{7}.
\newblock \DOIprefix\doi{10.3847/1538-4365/abb82d}, \href{http://arxiv.org/abs/1612.05243}{{\tt arXiv:1612.05243}}.
\bibitem[{{Fraser} and {Kavelaars}(2009)}]{kbo_size_distro}
\bibinfo{author}{{Fraser}, W.C.}, \bibinfo{author}{{Kavelaars}, J.J.}, \bibinfo{year}{2009}.
\newblock \bibinfo{title}{{The Size Distribution of Kuiper Belt Objects for D gsim 10 km}}.
\newblock \bibinfo{journal}{\aj} \bibinfo{volume}{137}, \bibinfo{pages}{72--82}.
\newblock \DOIprefix\doi{10.1088/0004-6256/137/1/72}, \href{http://arxiv.org/abs/0810.2296}{{\tt arXiv:0810.2296}}.
\bibitem[{{Gaia Collaboration} et~al.(2016){Gaia Collaboration}, {Prusti}, {de Bruijne}, {Brown}, {Vallenari}, {Babusiaux}, {Bailer-Jones}, {Bastian}, {Biermann}, {Evans}, {Eyer}, {Jansen}, {Jordi}, {Klioner}, {Lammers}, {Lindegren}, {Luri}, {Mignard}, {Milligan}, {Panem}, {Poinsignon}, {Pourbaix}, {Randich}, {Sarri}, {Sartoretti}, {Siddiqui}, {Soubiran}, {Valette}, {van Leeuwen}, {Walton}, {Aerts}, {Arenou}, {Cropper}, {Drimmel}, {H{\o}g}, {Katz}, {Lattanzi}, {O'Mullane}, {Grebel}, {Holland}, {Huc}, {Passot}, {Bramante}, {Cacciari}, {Casta{\~n}eda}, {Chaoul}, {Cheek}, {De Angeli}, {Fabricius}, {Guerra}, {Hern{\'a}ndez}, {Jean-Antoine-Piccolo}, {Masana}, {Messineo}, {Mowlavi}, {Nienartowicz}, {Ord{\'o}{\~n}ez-Blanco}, {Panuzzo}, {Portell}, {Richards}, {Riello}, {Seabroke}, {Tanga}, {Th{\'e}venin}, {Torra}, {Els}, {Gracia-Abril}, {Comoretto}, {Garcia-Reinaldos}, {Lock}, {Mercier}, {Altmann}, {Andrae}, {Astraatmadja}, {Bellas-Velidis}, {Benson}, {Berthier}, {Blomme}, {Busso}, {Carry}, {Cellino}, {Clementini},
  {Cowell}, {Creevey}, {Cuypers}, {Davidson}, {De Ridder}, {de Torres}, {Delchambre}, {Dell'Oro}, {Ducourant}, {Fr{\'e}mat}, {Garc{\'\i}a-Torres}, {Gosset}, {Halbwachs}, {Hambly}, {Harrison}, {Hauser}, {Hestroffer}, {Hodgkin}, {Huckle}, {Hutton}, {Jasniewicz}, {Jordan}, {Kontizas}, {Korn}, {Lanzafame}, {Manteiga}, {Moitinho}, {Muinonen}, {Osinde}, {Pancino}, {Pauwels}, {Petit}, {Recio-Blanco}, {Robin}, {Sarro}, {Siopis}, {Smith}, {Smith}, {Sozzetti}, {Thuillot}, {van Reeven}, {Viala}, {Abbas}, {Abreu Aramburu}, {Accart}, {Aguado}, {Allan}, {Allasia}, {Altavilla}, {{\'A}lvarez}, {Alves}, {Anderson}, {Andrei}, {Anglada Varela}, {Antiche}, {Antoja}, {Ant{\'o}n}, {Arcay}, {Atzei}, {Ayache}, {Bach}, {Baker}, {Balaguer-N{\'u}{\~n}ez}, {Barache}, {Barata}, {Barbier}, {Barblan}, {Baroni}, {Barrado y Navascu{\'e}s}, {Barros}, {Barstow}, {Becciani}, {Bellazzini}, {Bellei}, {Bello Garc{\'\i}a}, {Belokurov}, {Bendjoya}, {Berihuete}, {Bianchi}, {Bienaym{\'e}}, {Billebaud}, {Blagorodnova}, {Blanco-Cuaresma}, {Boch},
  {Bombrun}, {Borrachero}, {Bouquillon}, {Bourda}, {Bouy}, {Bragaglia}, {Breddels}, {Brouillet}, {Br{\"u}semeister}, {Bucciarelli}, {Budnik}, {Burgess}, {Burgon}, {Burlacu}, {Busonero}, {Buzzi}, {Caffau}, {Cambras}, {Campbell}, {Cancelliere}, {Cantat-Gaudin}, {Carlucci}, {Carrasco}, {Castellani}, {Charlot}, {Charnas}, {Charvet}, {Chassat}, {Chiavassa}, {Clotet}, {Cocozza}, {Collins}, {Collins} and {Costigan}}]{2016A&A...595A...1G}
\bibinfo{author}{{Gaia Collaboration}}, \bibinfo{author}{{Prusti}, T.}, \bibinfo{author}{{de Bruijne}, J.H.J.}, \bibinfo{author}{{Brown}, A.G.A.}, \bibinfo{author}{{Vallenari}, A.}, \bibinfo{author}{{Babusiaux}, C.}, \bibinfo{author}{{Bailer-Jones}, C.A.L.}, \bibinfo{author}{{Bastian}, U.}, \bibinfo{author}{{Biermann}, M.}, \bibinfo{author}{{Evans}, D.W.}, \bibinfo{author}{{Eyer}, L.}, \bibinfo{author}{{Jansen}, F.}, \bibinfo{author}{{Jordi}, C.}, \bibinfo{author}{{Klioner}, S.A.}, \bibinfo{author}{{Lammers}, U.}, \bibinfo{author}{{Lindegren}, L.}, \bibinfo{author}{{Luri}, X.}, \bibinfo{author}{{Mignard}, F.}, \bibinfo{author}{{Milligan}, D.J.}, \bibinfo{author}{{Panem}, C.}, \bibinfo{author}{{Poinsignon}, V.}, \bibinfo{author}{{Pourbaix}, D.}, \bibinfo{author}{{Randich}, S.}, \bibinfo{author}{{Sarri}, G.}, \bibinfo{author}{{Sartoretti}, P.}, \bibinfo{author}{{Siddiqui}, H.I.}, \bibinfo{author}{{Soubiran}, C.}, \bibinfo{author}{{Valette}, V.}, \bibinfo{author}{{van Leeuwen}, F.}, \bibinfo{author}{{Walton},
  N.A.}, \bibinfo{author}{{Aerts}, C.}, \bibinfo{author}{{Arenou}, F.}, \bibinfo{author}{{Cropper}, M.}, \bibinfo{author}{{Drimmel}, R.}, \bibinfo{author}{{H{\o}g}, E.}, \bibinfo{author}{{Katz}, D.}, \bibinfo{author}{{Lattanzi}, M.G.}, \bibinfo{author}{{O'Mullane}, W.}, \bibinfo{author}{{Grebel}, E.K.}, \bibinfo{author}{{Holland}, A.D.}, \bibinfo{author}{{Huc}, C.}, \bibinfo{author}{{Passot}, X.}, \bibinfo{author}{{Bramante}, L.}, \bibinfo{author}{{Cacciari}, C.}, \bibinfo{author}{{Casta{\~n}eda}, J.}, \bibinfo{author}{{Chaoul}, L.}, \bibinfo{author}{{Cheek}, N.}, \bibinfo{author}{{De Angeli}, F.}, \bibinfo{author}{{Fabricius}, C.}, \bibinfo{author}{{Guerra}, R.}, \bibinfo{author}{{Hern{\'a}ndez}, J.}, \bibinfo{author}{{Jean-Antoine-Piccolo}, A.}, \bibinfo{author}{{Masana}, E.}, \bibinfo{author}{{Messineo}, R.}, \bibinfo{author}{{Mowlavi}, N.}, \bibinfo{author}{{Nienartowicz}, K.}, \bibinfo{author}{{Ord{\'o}{\~n}ez-Blanco}, D.}, \bibinfo{author}{{Panuzzo}, P.}, \bibinfo{author}{{Portell}, J.},
  \bibinfo{author}{{Richards}, P.J.}, \bibinfo{author}{{Riello}, M.}, \bibinfo{author}{{Seabroke}, G.M.}, \bibinfo{author}{{Tanga}, P.}, \bibinfo{author}{{Th{\'e}venin}, F.}, \bibinfo{author}{{Torra}, J.}, \bibinfo{author}{{Els}, S.G.}, \bibinfo{author}{{Gracia-Abril}, G.}, \bibinfo{author}{{Comoretto}, G.}, \bibinfo{author}{{Garcia-Reinaldos}, M.}, \bibinfo{author}{{Lock}, T.}, \bibinfo{author}{{Mercier}, E.}, \bibinfo{author}{{Altmann}, M.}, \bibinfo{author}{{Andrae}, R.}, \bibinfo{author}{{Astraatmadja}, T.L.}, \bibinfo{author}{{Bellas-Velidis}, I.}, \bibinfo{author}{{Benson}, K.}, \bibinfo{author}{{Berthier}, J.}, \bibinfo{author}{{Blomme}, R.}, \bibinfo{author}{{Busso}, G.}, \bibinfo{author}{{Carry}, B.}, \bibinfo{author}{{Cellino}, A.}, \bibinfo{author}{{Clementini}, G.}, \bibinfo{author}{{Cowell}, S.}, \bibinfo{author}{{Creevey}, O.}, \bibinfo{author}{{Cuypers}, J.}, \bibinfo{author}{{Davidson}, M.}, \bibinfo{author}{{De Ridder}, J.}, \bibinfo{author}{{de Torres}, A.}, \bibinfo{author}{{Delchambre},
  L.}, \bibinfo{author}{{Dell'Oro}, A.}, \bibinfo{author}{{Ducourant}, C.}, \bibinfo{author}{{Fr{\'e}mat}, Y.}, \bibinfo{author}{{Garc{\'\i}a-Torres}, M.}, \bibinfo{author}{{Gosset}, E.}, \bibinfo{author}{{Halbwachs}, J.L.}, \bibinfo{author}{{Hambly}, N.C.}, \bibinfo{author}{{Harrison}, D.L.}, \bibinfo{author}{{Hauser}, M.}, \bibinfo{author}{{Hestroffer}, D.}, \bibinfo{author}{{Hodgkin}, S.T.}, \bibinfo{author}{{Huckle}, H.E.}, \bibinfo{author}{{Hutton}, A.}, \bibinfo{author}{{Jasniewicz}, G.}, \bibinfo{author}{{Jordan}, S.}, \bibinfo{author}{{Kontizas}, M.}, \bibinfo{author}{{Korn}, A.J.}, \bibinfo{author}{{Lanzafame}, A.C.}, \bibinfo{author}{{Manteiga}, M.}, \bibinfo{author}{{Moitinho}, A.}, \bibinfo{author}{{Muinonen}, K.}, \bibinfo{author}{{Osinde}, J.}, \bibinfo{author}{{Pancino}, E.}, \bibinfo{author}{{Pauwels}, T.}, \bibinfo{author}{{Petit}, J.M.}, \bibinfo{author}{{Recio-Blanco}, A.}, \bibinfo{author}{{Robin}, A.C.}, \bibinfo{author}{{Sarro}, L.M.}, \bibinfo{author}{{Siopis}, C.},
  \bibinfo{author}{{Smith}, M.}, \bibinfo{author}{{Smith}, K.W.}, \bibinfo{author}{{Sozzetti}, A.}, \bibinfo{author}{{Thuillot}, W.}, \bibinfo{author}{{van Reeven}, W.}, \bibinfo{author}{{Viala}, Y.}, \bibinfo{author}{{Abbas}, U.}, \bibinfo{author}{{Abreu Aramburu}, A.}, \bibinfo{author}{{Accart}, S.}, \bibinfo{author}{{Aguado}, J.J.}, \bibinfo{author}{{Allan}, P.M.}, \bibinfo{author}{{Allasia}, W.}, \bibinfo{author}{{Altavilla}, G.}, \bibinfo{author}{{{\'A}lvarez}, M.A.}, \bibinfo{author}{{Alves}, J.}, \bibinfo{author}{{Anderson}, R.I.}, \bibinfo{author}{{Andrei}, A.H.}, \bibinfo{author}{{Anglada Varela}, E.}, \bibinfo{author}{{Antiche}, E.}, \bibinfo{author}{{Antoja}, T.}, \bibinfo{author}{{Ant{\'o}n}, S.}, \bibinfo{author}{{Arcay}, B.}, \bibinfo{author}{{Atzei}, A.}, \bibinfo{author}{{Ayache}, L.}, \bibinfo{author}{{Bach}, N.}, \bibinfo{author}{{Baker}, S.G.}, \bibinfo{author}{{Balaguer-N{\'u}{\~n}ez}, L.}, \bibinfo{author}{{Barache}, C.}, \bibinfo{author}{{Barata}, C.}, \bibinfo{author}{{Barbier}, A.},
  \bibinfo{author}{{Barblan}, F.}, \bibinfo{author}{{Baroni}, M.}, \bibinfo{author}{{Barrado y Navascu{\'e}s}, D.}, \bibinfo{author}{{Barros}, M.}, \bibinfo{author}{{Barstow}, M.A.}, \bibinfo{author}{{Becciani}, U.}, \bibinfo{author}{{Bellazzini}, M.}, \bibinfo{author}{{Bellei}, G.}, \bibinfo{author}{{Bello Garc{\'\i}a}, A.}, \bibinfo{author}{{Belokurov}, V.}, \bibinfo{author}{{Bendjoya}, P.}, \bibinfo{author}{{Berihuete}, A.}, \bibinfo{author}{{Bianchi}, L.}, \bibinfo{author}{{Bienaym{\'e}}, O.}, \bibinfo{author}{{Billebaud}, F.}, \bibinfo{author}{{Blagorodnova}, N.}, \bibinfo{author}{{Blanco-Cuaresma}, S.}, \bibinfo{author}{{Boch}, T.}, \bibinfo{author}{{Bombrun}, A.}, \bibinfo{author}{{Borrachero}, R.}, \bibinfo{author}{{Bouquillon}, S.}, \bibinfo{author}{{Bourda}, G.}, \bibinfo{author}{{Bouy}, H.}, \bibinfo{author}{{Bragaglia}, A.}, \bibinfo{author}{{Breddels}, M.A.}, \bibinfo{author}{{Brouillet}, N.}, \bibinfo{author}{{Br{\"u}semeister}, T.}, \bibinfo{author}{{Bucciarelli}, B.},
  \bibinfo{author}{{Budnik}, F.}, \bibinfo{author}{{Burgess}, P.}, \bibinfo{author}{{Burgon}, R.}, \bibinfo{author}{{Burlacu}, A.}, \bibinfo{author}{{Busonero}, D.}, \bibinfo{author}{{Buzzi}, R.}, \bibinfo{author}{{Caffau}, E.}, \bibinfo{author}{{Cambras}, J.}, \bibinfo{author}{{Campbell}, H.}, \bibinfo{author}{{Cancelliere}, R.}, \bibinfo{author}{{Cantat-Gaudin}, T.}, \bibinfo{author}{{Carlucci}, T.}, \bibinfo{author}{{Carrasco}, J.M.}, \bibinfo{author}{{Castellani}, M.}, \bibinfo{author}{{Charlot}, P.}, \bibinfo{author}{{Charnas}, J.}, \bibinfo{author}{{Charvet}, P.}, \bibinfo{author}{{Chassat}, F.}, \bibinfo{author}{{Chiavassa}, A.}, \bibinfo{author}{{Clotet}, M.}, \bibinfo{author}{{Cocozza}, G.}, \bibinfo{author}{{Collins}, R.S.}, \bibinfo{author}{{Collins}, P.}, \bibinfo{author}{{Costigan}, G.}, \bibinfo{year}{2016}.
\newblock \bibinfo{title}{{The Gaia mission}}.
\newblock \bibinfo{journal}{\aap} \bibinfo{volume}{595}, \bibinfo{pages}{A1}.
\newblock \DOIprefix\doi{10.1051/0004-6361/201629272}, \href{http://arxiv.org/abs/1609.04153}{{\tt arXiv:1609.04153}}.
\bibitem[{{Giorgini}(2011)}]{2011jsrs.conf...87G}
\bibinfo{author}{{Giorgini}, J.}, \bibinfo{year}{2011}.
\newblock \bibinfo{title}{{Summary and status of the Horizons ephemeris system}}, in: \bibinfo{editor}{{Capitaine}, N.} (Ed.), \bibinfo{booktitle}{Journ{\'e}es Syst{\`e}mes de R{\'e}f{\'e}rence Spatio-temporels 2010}, pp. \bibinfo{pages}{87--87}.
\bibitem[{{Heinze} et~al.(2022){Heinze}, {Eggl}, {Juric}, {Moeyens}, {Jones}, {Sullivan} and {Bellm}}]{Heliolinc}
\bibinfo{author}{{Heinze}, A.}, \bibinfo{author}{{Eggl}, S.}, \bibinfo{author}{{Juric}, M.}, \bibinfo{author}{{Moeyens}, J.}, \bibinfo{author}{{Jones}, L.}, \bibinfo{author}{{Sullivan}, I.}, \bibinfo{author}{{Bellm}, E.}, \bibinfo{year}{2022}.
\newblock \bibinfo{title}{{Heliolinc3D: enabling asteroid discovery for the Legacy Survey of Space and Time (LSST)}}, in: \bibinfo{booktitle}{AAS/Division for Planetary Sciences Meeting Abstracts}, p. \bibinfo{pages}{504.04}.
\bibitem[{{Heinze} et~al.(2015){Heinze}, {Metchev} and {Trollo}}]{Heinze15}
\bibinfo{author}{{Heinze}, A.N.}, \bibinfo{author}{{Metchev}, S.}, \bibinfo{author}{{Trollo}, J.}, \bibinfo{year}{2015}.
\newblock \bibinfo{title}{{Digital Tracking Observations Can Discover Asteroids 10 Times Fainter Than Conventional Searches}}.
\newblock \bibinfo{journal}{\aj} \bibinfo{volume}{150}, \bibinfo{pages}{125}.
\newblock \DOIprefix\doi{10.1088/0004-6256/150/4/125}, \href{http://arxiv.org/abs/1508.01599}{{\tt arXiv:1508.01599}}.
\bibitem[{{Heinze} et~al.(2019){Heinze}, {Trollo} and {Metchev}}]{Heinze19}
\bibinfo{author}{{Heinze}, A.N.}, \bibinfo{author}{{Trollo}, J.}, \bibinfo{author}{{Metchev}, S.}, \bibinfo{year}{2019}.
\newblock \bibinfo{title}{{The Flux Distribution and Sky Density of 25th Magnitude Main Belt Asteroids}}.
\newblock \bibinfo{journal}{\aj} \bibinfo{volume}{158}, \bibinfo{pages}{232}.
\newblock \DOIprefix\doi{10.3847/1538-3881/ab48fa}, \href{http://arxiv.org/abs/1910.13015}{{\tt arXiv:1910.13015}}.
\bibitem[{Huang et~al.(2019)Huang, Cheng, Bapna, Firat, Chen, Chen, Lee, Ngiam, Le, Wu et~al.}]{huang2019gpipe}
\bibinfo{author}{Huang, Y.}, \bibinfo{author}{Cheng, Y.}, \bibinfo{author}{Bapna, A.}, \bibinfo{author}{Firat, O.}, \bibinfo{author}{Chen, D.}, \bibinfo{author}{Chen, M.}, \bibinfo{author}{Lee, H.}, \bibinfo{author}{Ngiam, J.}, \bibinfo{author}{Le, Q.V.}, \bibinfo{author}{Wu, Y.}, et~al., \bibinfo{year}{2019}.
\newblock \bibinfo{title}{Gpipe: Efficient training of giant neural networks using pipeline parallelism}.
\newblock \bibinfo{journal}{Advances in neural information processing systems} \bibinfo{volume}{32}.
\bibitem[{Ivezi\'c et~al.(2008)Ivezi\'c, Tyson, Acosta, Allsman, Anderson, Andrew, Angel, Axelrod, Barr, Becker et~al.}]{ivezic2008lsst}
\bibinfo{author}{Ivezi\'c, v.}, \bibinfo{author}{Tyson, J.A.}, \bibinfo{author}{Acosta, E.}, \bibinfo{author}{Allsman, R.}, \bibinfo{author}{Anderson, S.F.}, \bibinfo{author}{Andrew, J.}, \bibinfo{author}{Angel, J.R.P.}, \bibinfo{author}{Axelrod, T.S.}, \bibinfo{author}{Barr, J.D.}, \bibinfo{author}{Becker, A.C.}, et~al., \bibinfo{year}{2008}.
\newblock \bibinfo{title}{Lsst: from science drivers to reference design and anticipated data products} \href{http://arxiv.org/abs/0805.2366v4}{{\tt arXiv:0805.2366v4}}.
\bibitem[{{Juri{\'c}} et~al.(2015){Juri{\'c}}, {Kantor}, {Lim}, {Lupton}, {Dubois-Felsmann}, {Jenness}, {Axelrod}, {Aleksi{\'c}}, {Allsman}, {AlSayyad}, {Alt}, {Armstrong}, {Basney}, {Becker}, {Becla}, {Bickerton}, {Biswas}, {Bosch}, {Boutigny}, {Carrasco Kind}, {Ciardi}, {Connolly}, {Daniel}, {Daues}, {Economou}, {Chiang}, {Fausti}, {Fisher-Levine}, {Freemon}, {Gee}, {Gris}, {Hernandez}, {Hoblitt}, {Ivezi{\'c}}, {Jammes}, {Jevremovi{\'c}}, {Jones}, {Bryce Kalmbach}, {Kasliwal}, {Krughoff}, {Lang}, {Lurie}, {Lust}, {Mullally}, {MacArthur}, {Melchior}, {Moeyens}, {Nidever}, {Owen}, {Parejko}, {Peterson}, {Petravick}, {Pietrowicz}, {Price}, {Reiss}, {Shaw}, {Sick}, {Slater}, {Strauss}, {Sullivan}, {Swinbank}, {Van Dyk}, {Vuj{\v c}i{\'c}}, {Withers}, {Yoachim} and {LSST Project}}]{LSSTDM}
\bibinfo{author}{{Juri{\'c}}, M.}, \bibinfo{author}{{Kantor}, J.}, \bibinfo{author}{{Lim}, K.}, \bibinfo{author}{{Lupton}, R.H.}, \bibinfo{author}{{Dubois-Felsmann}, G.}, \bibinfo{author}{{Jenness}, T.}, \bibinfo{author}{{Axelrod}, T.S.}, \bibinfo{author}{{Aleksi{\'c}}, J.}, \bibinfo{author}{{Allsman}, R.A.}, \bibinfo{author}{{AlSayyad}, Y.}, \bibinfo{author}{{Alt}, J.}, \bibinfo{author}{{Armstrong}, R.}, \bibinfo{author}{{Basney}, J.}, \bibinfo{author}{{Becker}, A.C.}, \bibinfo{author}{{Becla}, J.}, \bibinfo{author}{{Bickerton}, S.J.}, \bibinfo{author}{{Biswas}, R.}, \bibinfo{author}{{Bosch}, J.}, \bibinfo{author}{{Boutigny}, D.}, \bibinfo{author}{{Carrasco Kind}, M.}, \bibinfo{author}{{Ciardi}, D.R.}, \bibinfo{author}{{Connolly}, A.J.}, \bibinfo{author}{{Daniel}, S.F.}, \bibinfo{author}{{Daues}, G.E.}, \bibinfo{author}{{Economou}, F.}, \bibinfo{author}{{Chiang}, H.F.}, \bibinfo{author}{{Fausti}, A.}, \bibinfo{author}{{Fisher-Levine}, M.}, \bibinfo{author}{{Freemon}, D.M.}, \bibinfo{author}{{Gee}, P.},
  \bibinfo{author}{{Gris}, P.}, \bibinfo{author}{{Hernandez}, F.}, \bibinfo{author}{{Hoblitt}, J.}, \bibinfo{author}{{Ivezi{\'c}}, {\v Z}.}, \bibinfo{author}{{Jammes}, F.}, \bibinfo{author}{{Jevremovi{\'c}}, D.}, \bibinfo{author}{{Jones}, R.L.}, \bibinfo{author}{{Bryce Kalmbach}, J.}, \bibinfo{author}{{Kasliwal}, V.P.}, \bibinfo{author}{{Krughoff}, K.S.}, \bibinfo{author}{{Lang}, D.}, \bibinfo{author}{{Lurie}, J.}, \bibinfo{author}{{Lust}, N.B.}, \bibinfo{author}{{Mullally}, F.}, \bibinfo{author}{{MacArthur}, L.A.}, \bibinfo{author}{{Melchior}, P.}, \bibinfo{author}{{Moeyens}, J.}, \bibinfo{author}{{Nidever}, D.L.}, \bibinfo{author}{{Owen}, R.}, \bibinfo{author}{{Parejko}, J.K.}, \bibinfo{author}{{Peterson}, J.M.}, \bibinfo{author}{{Petravick}, D.}, \bibinfo{author}{{Pietrowicz}, S.R.}, \bibinfo{author}{{Price}, P.A.}, \bibinfo{author}{{Reiss}, D.J.}, \bibinfo{author}{{Shaw}, R.A.}, \bibinfo{author}{{Sick}, J.}, \bibinfo{author}{{Slater}, C.T.}, \bibinfo{author}{{Strauss}, M.A.}, \bibinfo{author}{{Sullivan},
  I.S.}, \bibinfo{author}{{Swinbank}, J.D.}, \bibinfo{author}{{Van Dyk}, S.}, \bibinfo{author}{{Vuj{\v c}i{\'c}}, V.}, \bibinfo{author}{{Withers}, A.}, \bibinfo{author}{{Yoachim}, P.}, \bibinfo{author}{{LSST Project}, f.t.}, \bibinfo{year}{2015}.
\newblock \bibinfo{title}{{The LSST Data Management System}}.
\newblock \bibinfo{journal}{ArXiv e-prints} \href{http://arxiv.org/abs/1512.07914}{{\tt arXiv:1512.07914}}.
\bibitem[{Leclerc et~al.(2023)Leclerc, Ilyas, Engstrom, Park, Salman and Madry}]{leclerc2023ffcv_bottlenecks}
\bibinfo{author}{Leclerc, G.}, \bibinfo{author}{Ilyas, A.}, \bibinfo{author}{Engstrom, L.}, \bibinfo{author}{Park, S.M.}, \bibinfo{author}{Salman, H.}, \bibinfo{author}{Madry, A.}, \bibinfo{year}{2023}.
\newblock \bibinfo{title}{Ffcv: Accelerating training by removing data bottlenecks}, in: \bibinfo{booktitle}{Proceedings of the IEEE/CVF Conference on Computer Vision and Pattern Recognition (CVPR)}, pp. \bibinfo{pages}{12011--12020}.
\bibitem[{{Lifset} et~al.(2021){Lifset}, {Golovich}, {Green}, {Armstrong} and {Yeager}}]{Lifset21}
\bibinfo{author}{{Lifset}, N.}, \bibinfo{author}{{Golovich}, N.}, \bibinfo{author}{{Green}, E.}, \bibinfo{author}{{Armstrong}, R.}, \bibinfo{author}{{Yeager}, T.}, \bibinfo{year}{2021}.
\newblock \bibinfo{title}{{A Search for L4 Earth Trojan Asteroids Using a Novel Track-before-detect Multiepoch Pipeline}}.
\newblock \bibinfo{journal}{\aj} \bibinfo{volume}{161}, \bibinfo{pages}{282}.
\newblock \DOIprefix\doi{10.3847/1538-3881/abf7af}, \href{http://arxiv.org/abs/2102.09059}{{\tt arXiv:2102.09059}}.
\bibitem[{{Masci} et~al.(2023){Masci}, {Laher}, {Rusholme}, {Shupe}, {Paladini}, {Groom}, {Wold}, {Miller} and {Drake}}]{2023arXiv230516279M}
\bibinfo{author}{{Masci}, F.J.}, \bibinfo{author}{{Laher}, R.R.}, \bibinfo{author}{{Rusholme}, B.}, \bibinfo{author}{{Shupe}, D.}, \bibinfo{author}{{Paladini}, R.}, \bibinfo{author}{{Groom}, S.}, \bibinfo{author}{{Wold}, A.}, \bibinfo{author}{{Miller}, A.A.}, \bibinfo{author}{{Drake}, A.}, \bibinfo{year}{2023}.
\newblock \bibinfo{title}{{A New Forced Photometry Service for the Zwicky Transient Facility}}.
\newblock \bibinfo{journal}{arXiv e-prints} , \bibinfo{pages}{arXiv:2305.16279}\DOIprefix\doi{10.48550/arXiv.2305.16279}, \href{http://arxiv.org/abs/2305.16279}{{\tt arXiv:2305.16279}}.
\bibitem[{{Masci} et~al.(2019){Masci}, {Laher}, {Rusholme}, {Shupe}, {Groom}, {Surace}, {Jackson}, {Monkewitz}, {Beck}, {Flynn}, {Terek}, {Landry}, {Hacopians}, {Desai}, {Howell}, {Brooke}, {Imel}, {Wachter}, {Ye}, {Lin}, {Cenko}, {Cunningham}, {Rebbapragada}, {Bue}, {Miller}, {Mahabal}, {Bellm}, {Patterson}, {Juri{\'c}}, {Golkhou}, {Ofek}, {Walters}, {Graham}, {Kasliwal}, {Dekany}, {Kupfer}, {Burdge}, {Cannella}, {Barlow}, {Van Sistine}, {Giomi}, {Fremling}, {Blagorodnova}, {Levitan}, {Riddle}, {Smith}, {Helou}, {Prince} and {Kulkarni}}]{ztf2}
\bibinfo{author}{{Masci}, F.J.}, \bibinfo{author}{{Laher}, R.R.}, \bibinfo{author}{{Rusholme}, B.}, \bibinfo{author}{{Shupe}, D.L.}, \bibinfo{author}{{Groom}, S.}, \bibinfo{author}{{Surace}, J.}, \bibinfo{author}{{Jackson}, E.}, \bibinfo{author}{{Monkewitz}, S.}, \bibinfo{author}{{Beck}, R.}, \bibinfo{author}{{Flynn}, D.}, \bibinfo{author}{{Terek}, S.}, \bibinfo{author}{{Landry}, W.}, \bibinfo{author}{{Hacopians}, E.}, \bibinfo{author}{{Desai}, V.}, \bibinfo{author}{{Howell}, J.}, \bibinfo{author}{{Brooke}, T.}, \bibinfo{author}{{Imel}, D.}, \bibinfo{author}{{Wachter}, S.}, \bibinfo{author}{{Ye}, Q.Z.}, \bibinfo{author}{{Lin}, H.W.}, \bibinfo{author}{{Cenko}, S.B.}, \bibinfo{author}{{Cunningham}, V.}, \bibinfo{author}{{Rebbapragada}, U.}, \bibinfo{author}{{Bue}, B.}, \bibinfo{author}{{Miller}, A.A.}, \bibinfo{author}{{Mahabal}, A.}, \bibinfo{author}{{Bellm}, E.C.}, \bibinfo{author}{{Patterson}, M.T.}, \bibinfo{author}{{Juri{\'c}}, M.}, \bibinfo{author}{{Golkhou}, V.Z.}, \bibinfo{author}{{Ofek}, E.O.},
  \bibinfo{author}{{Walters}, R.}, \bibinfo{author}{{Graham}, M.}, \bibinfo{author}{{Kasliwal}, M.M.}, \bibinfo{author}{{Dekany}, R.G.}, \bibinfo{author}{{Kupfer}, T.}, \bibinfo{author}{{Burdge}, K.}, \bibinfo{author}{{Cannella}, C.B.}, \bibinfo{author}{{Barlow}, T.}, \bibinfo{author}{{Van Sistine}, A.}, \bibinfo{author}{{Giomi}, M.}, \bibinfo{author}{{Fremling}, C.}, \bibinfo{author}{{Blagorodnova}, N.}, \bibinfo{author}{{Levitan}, D.}, \bibinfo{author}{{Riddle}, R.}, \bibinfo{author}{{Smith}, R.M.}, \bibinfo{author}{{Helou}, G.}, \bibinfo{author}{{Prince}, T.A.}, \bibinfo{author}{{Kulkarni}, S.R.}, \bibinfo{year}{2019}.
\newblock \bibinfo{title}{{The Zwicky Transient Facility: Data Processing, Products, and Archive}}.
\newblock \bibinfo{journal}{\pasp} \bibinfo{volume}{131}, \bibinfo{pages}{018003}.
\newblock \DOIprefix\doi{10.1088/1538-3873/aae8ac}, \href{http://arxiv.org/abs/1902.01872}{{\tt arXiv:1902.01872}}.
\bibitem[{Mohan et~al.(2020)Mohan, Phanishayee, Raniwala and Chidambaram}]{mohan2020DNN_data_stalls}
\bibinfo{author}{Mohan, J.}, \bibinfo{author}{Phanishayee, A.}, \bibinfo{author}{Raniwala, A.}, \bibinfo{author}{Chidambaram, V.}, \bibinfo{year}{2020}.
\newblock \bibinfo{title}{Analyzing and mitigating data stalls in dnn training}.
\newblock \bibinfo{journal}{arXiv preprint arXiv:2007.06775} .
\bibitem[{{Morbidelli} et~al.(2021){Morbidelli}, {Nesvorny}, {Bottke} and {Marchi}}]{kbo_size_distro2}
\bibinfo{author}{{Morbidelli}, A.}, \bibinfo{author}{{Nesvorny}, D.}, \bibinfo{author}{{Bottke}, W.F.}, \bibinfo{author}{{Marchi}, S.}, \bibinfo{year}{2021}.
\newblock \bibinfo{title}{{A re-assessment of the Kuiper belt size distribution for sub-kilometer objects, revealing collisional equilibrium at small sizes}}.
\newblock \bibinfo{journal}{\icarus} \bibinfo{volume}{356}, \bibinfo{pages}{114256}.
\newblock \DOIprefix\doi{10.1016/j.icarus.2020.114256}, \href{http://arxiv.org/abs/2012.03823}{{\tt arXiv:2012.03823}}.
\bibitem[{{Park} et~al.(2021){Park}, {Folkner}, {Williams} and {Boggs}}]{2021AJ....161..105P}
\bibinfo{author}{{Park}, R.S.}, \bibinfo{author}{{Folkner}, W.M.}, \bibinfo{author}{{Williams}, J.G.}, \bibinfo{author}{{Boggs}, D.H.}, \bibinfo{year}{2021}.
\newblock \bibinfo{title}{{The JPL Planetary and Lunar Ephemerides DE440 and DE441}}.
\newblock \bibinfo{journal}{\aj} \bibinfo{volume}{161}, \bibinfo{pages}{105}.
\newblock \DOIprefix\doi{10.3847/1538-3881/abd414}.
\bibitem[{{Pence} et~al.(2011){Pence}, {Seaman} and {White}}]{fpack}
\bibinfo{author}{{Pence}, W.}, \bibinfo{author}{{Seaman}, R.}, \bibinfo{author}{{White}, R.}, \bibinfo{year}{2011}.
\newblock \bibinfo{title}{{Fpack and Funpack User's Guide: FITS Image Compression Utilities}}.
\newblock \bibinfo{journal}{arXiv e-prints} , \bibinfo{pages}{arXiv:1112.2671}\DOIprefix\doi{10.48550/arXiv.1112.2671}, \href{http://arxiv.org/abs/1112.2671}{{\tt arXiv:1112.2671}}.
\bibitem[{Schlafly et~al.(2023)Schlafly, Yeager, Pruett, Schneider, Ebert, Merl, Lifset, Armstrong, Dawson, Meyers, Perloff and Golovich}]{ssapy}
\bibinfo{author}{Schlafly, E.}, \bibinfo{author}{Yeager, T.}, \bibinfo{author}{Pruett, K.}, \bibinfo{author}{Schneider, M.}, \bibinfo{author}{Ebert, J.}, \bibinfo{author}{Merl, D.}, \bibinfo{author}{Lifset, N.}, \bibinfo{author}{Armstrong, R.}, \bibinfo{author}{Dawson, W.}, \bibinfo{author}{Meyers, J.}, \bibinfo{author}{Perloff, A.}, \bibinfo{author}{Golovich, N.}, \bibinfo{year}{2023}.
\newblock \bibinfo{title}{Space situational awareness for python}.
\newblock \bibinfo{howpublished}{[Computer Software] \url{https://doi.org/10.11578/dc.20240417.1}}.
\newblock \URLprefix \url{https://doi.org/10.11578/dc.20240417.1}, \DOIprefix\doi{10.11578/dc.20240417.1}.
\bibitem[{{Shao} et~al.(2014){Shao}, {Nemati}, {Zhai}, {Turyshev}, {Sandhu}, {Hallinan} and {Harding}}]{Shao14}
\bibinfo{author}{{Shao}, M.}, \bibinfo{author}{{Nemati}, B.}, \bibinfo{author}{{Zhai}, C.}, \bibinfo{author}{{Turyshev}, S.G.}, \bibinfo{author}{{Sandhu}, J.}, \bibinfo{author}{{Hallinan}, G.}, \bibinfo{author}{{Harding}, L.K.}, \bibinfo{year}{2014}.
\newblock \bibinfo{title}{{Finding Very Small Near-Earth Asteroids using Synthetic Tracking}}.
\newblock \bibinfo{journal}{\apj} \bibinfo{volume}{782}, \bibinfo{pages}{1}.
\newblock \DOIprefix\doi{10.1088/0004-637X/782/1/1}, \href{http://arxiv.org/abs/1309.3248}{{\tt arXiv:1309.3248}}.
\bibitem[{Steil et~al.(2023)Steil, Reza, Priest and Pearce}]{steil2023ygm}
\bibinfo{author}{Steil, T.}, \bibinfo{author}{Reza, T.}, \bibinfo{author}{Priest, B.}, \bibinfo{author}{Pearce, R.}, \bibinfo{year}{2023}.
\newblock \bibinfo{title}{Embracing irregular parallelism in hpc with ygm}, in: \bibinfo{booktitle}{Proceedings of the International Conference for High Performance Computing, Networking, Storage and Analysis}, \bibinfo{publisher}{Association for Computing Machinery}, \bibinfo{address}{New York, NY, USA}.
\newblock \URLprefix \url{https://doi.org/10.1145/3581784.3607103}, \DOIprefix\doi{10.1145/3581784.3607103}.
\bibitem[{Thomas(2024)}]{el-capitan}
\bibinfo{author}{Thomas, J.}, \bibinfo{year}{2024}.
\newblock \bibinfo{title}{{Lawrence Livermore National Laboratory’s El Capitan} verified as world's fastest supercomputer}.
\newblock \bibinfo{howpublished}{\url{https://www.llnl.gov/article/52061/lawrence-livermore-national-laboratorys-el-capitan-verified-worlds-fastest-supercomputer}}.
\bibitem[{{Tonry} et~al.(2018){Tonry}, {Denneau}, {Heinze}, {Stalder}, {Smith}, {Smartt}, {Stubbs}, {Weiland} and {Rest}}]{atlas}
\bibinfo{author}{{Tonry}, J.L.}, \bibinfo{author}{{Denneau}, L.}, \bibinfo{author}{{Heinze}, A.N.}, \bibinfo{author}{{Stalder}, B.}, \bibinfo{author}{{Smith}, K.W.}, \bibinfo{author}{{Smartt}, S.J.}, \bibinfo{author}{{Stubbs}, C.W.}, \bibinfo{author}{{Weiland}, H.J.}, \bibinfo{author}{{Rest}, A.}, \bibinfo{year}{2018}.
\newblock \bibinfo{title}{{ATLAS: A High-cadence All-sky Survey System}}.
\newblock \bibinfo{journal}{\pasp} \bibinfo{volume}{130}, \bibinfo{pages}{064505}.
\newblock \DOIprefix\doi{10.1088/1538-3873/aabadf}, \href{http://arxiv.org/abs/1802.00879}{{\tt arXiv:1802.00879}}.
\bibitem[{{Tyson} et~al.(1992){Tyson}, {Guhathakurta}, {Bernstein} and {Hut}}]{Tyson92}
\bibinfo{author}{{Tyson}, J.A.}, \bibinfo{author}{{Guhathakurta}, P.}, \bibinfo{author}{{Bernstein}, G.M.}, \bibinfo{author}{{Hut}, P.}, \bibinfo{year}{1992}.
\newblock \bibinfo{title}{{Limits on the Surface Density of Faint Kuiper Belt Objects}}, in: \bibinfo{booktitle}{American Astronomical Society Meeting Abstracts}, p. \bibinfo{pages}{06.10}.
\bibitem[{Van~Essen et~al.(2012)Van~Essen, Hsieh, Ames and Gokhale}]{vanessen2012di-mmap}
\bibinfo{author}{Van~Essen, B.}, \bibinfo{author}{Hsieh, H.}, \bibinfo{author}{Ames, S.}, \bibinfo{author}{Gokhale, M.}, \bibinfo{year}{2012}.
\newblock \bibinfo{title}{Di-mmap: A high performance memory-map runtime for data-intensive applications}, in: \bibinfo{booktitle}{2012 SC Companion: High Performance Computing, Networking Storage and Analysis}, pp. \bibinfo{pages}{731--735}.
\newblock \DOIprefix\doi{10.1109/SC.Companion.2012.99}.
\bibitem[{Wang et~al.(2011)Wang, Yang, Du, Chen, Yi and Xu}]{wang2011optimizing}
\bibinfo{author}{Wang, F.}, \bibinfo{author}{Yang, C.Q.}, \bibinfo{author}{Du, Y.F.}, \bibinfo{author}{Chen, J.}, \bibinfo{author}{Yi, H.Z.}, \bibinfo{author}{Xu, W.X.}, \bibinfo{year}{2011}.
\newblock \bibinfo{title}{Optimizing linpack benchmark on gpu-accelerated petascale supercomputer}.
\newblock \bibinfo{journal}{Journal of Computer Science and Technology} \bibinfo{volume}{26}, \bibinfo{pages}{854--865}.
\bibitem[{{Wells} et~al.(1981){Wells}, {Greisen} and {Harten}}]{fits}
\bibinfo{author}{{Wells}, D.C.}, \bibinfo{author}{{Greisen}, E.W.}, \bibinfo{author}{{Harten}, R.H.}, \bibinfo{year}{1981}.
\newblock \bibinfo{title}{{FITS - a Flexible Image Transport System}}.
\newblock \bibinfo{journal}{\aaps} \bibinfo{volume}{44}, \bibinfo{pages}{363}.
\bibitem[{{Whidden} et~al.(2019){Whidden}, {Bryce Kalmbach}, {Connolly}, {Jones}, {Smotherman}, {Bektesevic}, {Slater}, {Becker}, {Ivezi{\'c}}, {Juri{\'c}}, {Bolin}, {Moeyens}, {F{\"o}rster} and {Golkhou}}]{WhiddenGPU}
\bibinfo{author}{{Whidden}, P.J.}, \bibinfo{author}{{Bryce Kalmbach}, J.}, \bibinfo{author}{{Connolly}, A.J.}, \bibinfo{author}{{Jones}, R.L.}, \bibinfo{author}{{Smotherman}, H.}, \bibinfo{author}{{Bektesevic}, D.}, \bibinfo{author}{{Slater}, C.}, \bibinfo{author}{{Becker}, A.C.}, \bibinfo{author}{{Ivezi{\'c}}, {\v{Z}}.}, \bibinfo{author}{{Juri{\'c}}, M.}, \bibinfo{author}{{Bolin}, B.}, \bibinfo{author}{{Moeyens}, J.}, \bibinfo{author}{{F{\"o}rster}, F.}, \bibinfo{author}{{Golkhou}, V.Z.}, \bibinfo{year}{2019}.
\newblock \bibinfo{title}{{Fast Algorithms for Slow Moving Asteroids: Constraints on the Distribution of Kuiper Belt Objects}}.
\newblock \bibinfo{journal}{\aj} \bibinfo{volume}{157}, \bibinfo{pages}{119}.
\newblock \DOIprefix\doi{10.3847/1538-3881/aafd2d}, \href{http://arxiv.org/abs/1901.02492}{{\tt arXiv:1901.02492}}.
\bibitem[{{Yeager} et~al.(2023){Yeager}, {Pruett} and {Schneider}}]{yeager2023cislunar}
\bibinfo{author}{{Yeager}, T.}, \bibinfo{author}{{Pruett}, K.}, \bibinfo{author}{{Schneider}, M.}, \bibinfo{year}{2023}.
\newblock \bibinfo{title}{{Long-term N-body Stability in Cislunar Space}}, in: \bibinfo{editor}{{Ryan}, S.} (Ed.), \bibinfo{booktitle}{Proceedings of the Advanced Maui Optical and Space Surveillance (AMOS) Technologies Conference}, p. \bibinfo{pages}{208}.
\bibitem[{{Zhai} et~al.(2014){Zhai}, {Shao}, {Nemati}, {Werne}, {Zhou}, {Turyshev}, {Sandhu}, {Hallinan} and {Harding}}]{Zhai14}
\bibinfo{author}{{Zhai}, C.}, \bibinfo{author}{{Shao}, M.}, \bibinfo{author}{{Nemati}, B.}, \bibinfo{author}{{Werne}, T.}, \bibinfo{author}{{Zhou}, H.}, \bibinfo{author}{{Turyshev}, S.G.}, \bibinfo{author}{{Sandhu}, J.}, \bibinfo{author}{{Hallinan}, G.}, \bibinfo{author}{{Harding}, L.K.}, \bibinfo{year}{2014}.
\newblock \bibinfo{title}{{Detection of a Faint Fast-moving Near-Earth Asteroid Using the Synthetic Tracking Technique}}.
\newblock \bibinfo{journal}{\apj} \bibinfo{volume}{792}, \bibinfo{pages}{60}.
\newblock \DOIprefix\doi{10.1088/0004-637X/792/1/60}, \href{http://arxiv.org/abs/1403.4353}{{\tt arXiv:1403.4353}}.
\bibitem[{{Zhai} et~al.(2018){Zhai}, {Shao}, {Saini}, {Sandhu}, {Owen}, {Choi}, {Werne}, {Ely}, {Lazio}, {Martin-Mur}, {Preston}, {Turyshev}, {Mitchell}, {Nazli}, {Cui} and {Mochama}}]{Zhai18}
\bibinfo{author}{{Zhai}, C.}, \bibinfo{author}{{Shao}, M.}, \bibinfo{author}{{Saini}, N.S.}, \bibinfo{author}{{Sandhu}, J.S.}, \bibinfo{author}{{Owen}, W.M.}, \bibinfo{author}{{Choi}, P.}, \bibinfo{author}{{Werne}, T.A.}, \bibinfo{author}{{Ely}, T.A.}, \bibinfo{author}{{Lazio}, J.}, \bibinfo{author}{{Martin-Mur}, T.J.}, \bibinfo{author}{{Preston}, R.A.}, \bibinfo{author}{{Turyshev}, S.G.}, \bibinfo{author}{{Mitchell}, A.W.}, \bibinfo{author}{{Nazli}, K.}, \bibinfo{author}{{Cui}, I.}, \bibinfo{author}{{Mochama}, R.M.}, \bibinfo{year}{2018}.
\newblock \bibinfo{title}{{Accurate Ground-based Near-Earth-Asteroid Astrometry Using Synthetic Tracking}}.
\newblock \bibinfo{journal}{\aj} \bibinfo{volume}{156}, \bibinfo{pages}{65}.
\newblock \DOIprefix\doi{10.3847/1538-3881/aacb28}, \href{http://arxiv.org/abs/1805.01107}{{\tt arXiv:1805.01107}}.
\bibitem[{{Zhai} et~al.(2020){Zhai}, {Ye}, {Shao}, {Trahan}, {Saini}, {Shen}, {Prince}, {Bellm}, {Graham}, {Helou}, {Kulkarni}, {Kupfer}, {Laher}, {Mahabal}, {Masci}, {Rusholme}, {Rosnet} and {Shupe}}]{Zhai20}
\bibinfo{author}{{Zhai}, C.}, \bibinfo{author}{{Ye}, Q.}, \bibinfo{author}{{Shao}, M.}, \bibinfo{author}{{Trahan}, R.}, \bibinfo{author}{{Saini}, N.S.}, \bibinfo{author}{{Shen}, J.}, \bibinfo{author}{{Prince}, T.A.}, \bibinfo{author}{{Bellm}, E.C.}, \bibinfo{author}{{Graham}, M.J.}, \bibinfo{author}{{Helou}, G.}, \bibinfo{author}{{Kulkarni}, S.R.}, \bibinfo{author}{{Kupfer}, T.}, \bibinfo{author}{{Laher}, R.R.}, \bibinfo{author}{{Mahabal}, A.}, \bibinfo{author}{{Masci}, F.J.}, \bibinfo{author}{{Rusholme}, B.}, \bibinfo{author}{{Rosnet}, P.}, \bibinfo{author}{{Shupe}, D.L.}, \bibinfo{year}{2020}.
\newblock \bibinfo{title}{{Synthetic Tracking Using ZTF Deep Drilling Data Sets}}.
\newblock \bibinfo{journal}{\pasp} \bibinfo{volume}{132}, \bibinfo{pages}{064502}.
\newblock \DOIprefix\doi{10.1088/1538-3873/ab828b}, \href{http://arxiv.org/abs/1907.11299}{{\tt arXiv:1907.11299}}.

\end{thebibliography}

\end{document}